%% file: submission_taslp14F.tex
\documentclass[10pt,twocolumn,twoside]{IEEEtran}
\usepackage{graphicx}
\usepackage{amsmath}
\usepackage{amsfonts}
\usepackage{amssymb}
\usepackage{amsthm}
\usepackage{amsopn}
\usepackage{xspace}
\usepackage{array}
\usepackage{epsfig}
\usepackage{subfigure}
\usepackage{math}
\usepackage{float}
\usepackage{color}
\usepackage{url}
\usepackage{cleveref}
\crefformat{footnote}{#2\footnotemark[#1]#3}

\newtheorem{theorem}{Theorem}

\begin{document}
%
%
%
%
\title{Co-Localization of Audio Sources in Images Using Binaural Features and Locally-Linear Regression}
\author{
\IEEEauthorblockN{{Antoine Deleforge}\IEEEauthorrefmark{1}} \and
\IEEEauthorblockN{{Radu Horaud}\IEEEauthorrefmark{1}} \and
\IEEEauthorblockN{{Yoav Y.~Schechner}\IEEEauthorrefmark{3}}  \and
\IEEEauthorblockN{{Laurent Girin}\IEEEauthorrefmark{1}\IEEEauthorrefmark{2}} 
\\
\IEEEauthorblockA{\IEEEauthorrefmark{1} INRIA Grenoble Rh\^one-Alpes, Montbonnot Saint-Martin, France}
\\
\IEEEauthorblockA{\IEEEauthorrefmark{2} Univ. Grenoble Alpes, GIPSA-Lab, France} 
\\
\IEEEauthorblockA{\IEEEauthorrefmark{3} Dept. Electrical Eng., Technion-Israel Inst. of Technology, Haifa, Israel} 
\thanks{A. Deleforge, R. Horaud and L. Girin acknowledge support from 
the European Research Council through the ERC Advanced Grant VHIA \#340113.}
\thanks{Y. Y. Schechner is supported by the Technion Autonomous Systems Program (TASP) and Israel Science Foundation (ISF) Grant 1467/12. It was partially conducted in the Ollendorff Minerva Center, funded through the BMBF. YYS is a Landau Fellow supported by the Taub Foundation. }}
%
%

\markboth{Transactions on Audio, Speech, and Language Processing, vol.~23, no.~4, April~2015}%
{Deleforge \MakeLowercase{\textit{et al.}}: Co-Localization of Audio Sources in Images Using Binaural Cues and Locally-Linear Regression}
%



\maketitle

\input{abstractF}


%

\input{introductionF}

\input{spaceF}

\input{gllimF}

\input{resultsF}

\input{conclusionF}

\section{Acknowledgements}
The authors are grateful to Israel-Dejene Gebru and Vincent Drouard for their precious help with data collection and preparation. They also warmly thank the anonymous reviewers for their dedicated reviews with serious and highly valuable comments and suggestions.

\ifCLASSOPTIONcaptionsoff
  \newpage
\fi



%

\bibliographystyle{IEEEtran}

%

\input{biographies}





\end{document}

%% file: abstractF.tex
\begin{abstract}

This paper addresses the problem of localizing audio sources using binaural measurements. We propose a supervised formulation that simultaneously localizes multiple sources at different locations. The approach is intrinsically efficient because, contrary to prior work, it relies neither on source separation, nor on monaural segregation. The method starts with a training stage that establishes a locally-linear Gaussian regression model between the directional coordinates of all the sources and the auditory features extracted 
from binaural measurements. While fixed-length wide-spectrum sounds (white noise) are used for training to reliably estimate the model parameters, we show that the testing (localization) can be extended to variable-length sparse-spectrum sounds (such as speech), thus enabling a wide range of realistic applications. Indeed, we demonstrate that the method can be used for audio-visual fusion, namely to map speech signals onto images and hence to spatially align the audio and visual modalities, thus enabling to discriminate between speaking and non-speaking faces. We release a novel corpus of real-room recordings that allow quantitative evaluation of the co-localization method in the presence of one or two sound sources. Experiments demonstrate increased accuracy and speed relative to several state-of-the-art methods.


\end{abstract}

\begin{IEEEkeywords}
  Sound-source localization, binaural hearing, supervised learning, mixture model, regression, audio-visual fusion.
  \end{IEEEkeywords}

%% file: introductionF.tex
\section{Introduction}
\label{sec:introduction}
\IEEEPARstart{W}{e} address the problem of localizing one or several sound sources from recordings gathered with two microphones plugged into the ears of an acoustic dummy head. This problem is of particular interest in the context of a humanoid robot analyzing auditory scenes to better interact with its environment, \textit{e.g.} \cite{cech2013active}.  The shape and morphology of such a \textit{binaural} setup induce filtering effects,  and hence discrepancies in both intensity-level and phase, at each frequency band, between the two microphone signals.
These discrepancies are the \textit{interaural level difference} (ILD) and the \textit{interaural time difference} (ITD) or equivalently the \textit{interaural phase difference} (IPD). The ILD and IPD values across all frequencies are referred as {\em binaural features}.

For a single spatially-narrow emitter, the ILD and IPD depend on the emitter's position relative to the head, namely the 2D directional vector formed by 
azimuth and elevation. 
Binaural features have hence been used for single sound source localization (single-SSL), \textit{e.g.}, \cite{datum1996artificial,WillertEggertAdamyStahlKorner2006,kulaib20092d,HornsteinLopesSantosVictorLacerda2006,LuCooke2010,Raspaud2010,talmon2011supervised,DeleforgeHoraud12,Luo2014,Keyrouz2014}.
Matters are more complex when multiple sound sources, emitting from different directions, are simultaneously active. The sources mix at each microphone and the binaural features not only depend on the unknown emitting directions but also on the unknown emitted spectra. 

A common approximation assumes that any observed time-frequency (TF) point that has significant acoustic power is dominated by just a single source.
This assumption, referred to as \textit{W-disjoint orthogonality} (WDO)~\cite{YilmazRickard04} simplifies the analysis: The binaural information at a TF point is simply related to the direction of a single source. 
WDO has been shown to be valid to some extent in the case of mixtures of speech signals, though it may have limitations in dense cocktail party scenarios.

State-of-the-art multiple-SSL techniques strongly rely on WDO to \textit{spatially group} binaural features, \textit{i.e.}, to assign a given TF point to a single source~\cite{Aarabi02,RomanWangBrown2003,YilmazRickard04,RomanWang2008,MandelWeissEllis10,LeePark2010,WoodruffWang2012,DeleforgeForbesHoraud13}. Some of these methods perform the grouping by selecting peaks in histograms of ITDs, accumulated over frequency channels~\cite{Aarabi02,YilmazRickard04,RomanWang2008,LeePark2010}. Other methods iteratively alternate between separation and localization \cite{MandelWeissEllis10,DeleforgeForbesHoraud13}. They require expectation-maximization (EM) inference at runtime, which is computationally intensive.
The WDO assumption can also be combined with monaural segregation. For example, in \cite{RomanWangBrown2003,WoodruffWang2010,WoodruffWang2012} azimuth is estimated from only those TF points at which a single source is thought to be dominant based on voiced and unvoiced speech cues. These monaural cues are then combined with a statistical model of ILD/ITD distribution that takes into account interfering sources, reverberation or background noise. 

The vast majority of the above-mentioned techniques limit single- and multiple-SSL to 1D localization, namely along the frontal azimuth direction and are based on a simplified sound propagation model. Moreover, these methods attempt to extract localization information based on a physical model that must be somehow explicitly identified and inverted, \textit{e.g.}, the head-related transfer functions (HRTF) of the system. 

We propose a method that directly localizes either a single or several 
  sources simultaneously, on the following grounds:
\begin{itemize}
\item it doesn't rely on the WDO assumption, on source separation, or on monaural segregation;
\item it is based on learning a regression model that implicitly encodes the HRTF using training data;
\item it can use single-source data to train a multiple-source localization model;
\item it outperforms competing methods in terms of robustness, accuracy, and speed, and
\item it can be used to map sound sources onto images.
\end{itemize}



 \begin{figure*}[h!t!]
  \centering
  \begin{tabular}{ccc}
   \includegraphics[height = 0.25\linewidth]{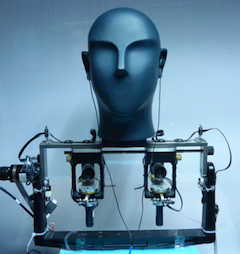} & 
     \includegraphics[height = 0.25\linewidth]{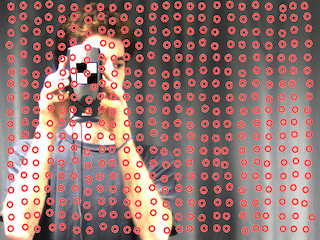} &
     \includegraphics[height = 0.25\linewidth]{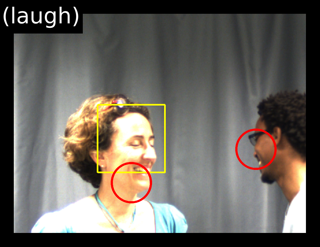}\\
      \end{tabular}
  \caption{\label{fig:principle}  \textit{Left: Binaural recordings} and associated sound-source directions are simultaneously gathered with two microphones plugged into the ears of a dummy head and a camera placed under the head (only one of the two cameras is used in this work). The dummy head induces non-isotropic filtering effects responsible for 2D sound localization.  \textit{Middle: Training data} are obtained as follows. A sound-direction-to-binaural-feature (low-to-high dimensional) regression is learned using an audio-visual target composed of a loud speaker and a visual marker. The loud-speaker that emits \textbf{fixed-length full-spectrum sounds} is moved in front of the dummy-head/camera device and for each loud-speaker location, both the emitted sound and the image location of the visual marker are recorded. The tiny red circles correspond to the 432 locations of the loud-speaker used for training. \textit{Right: Multiple sound-source localization}. Based on the parameters of the trained regression, the sound directions (or equivalently, image locations) are estimated from a \textbf{variable-length sparse spectrogram} (large red circles) in near real-time. The yellow square corresponds to the result of a face detector \cite{viola2004robust} which can only deal with frontal views.}
\end{figure*}

\subsection{Related Work and Contributions}
\label{subsection:related_work}
To overcome the need of a complex explicit sound propagation model, a number of
\textit{supervised} approaches to SSL have been recently proposed. These methods use either artificial neural networks \cite{datum1996artificial}, manifold learning \cite{talmon2011supervised,DeleforgeHoraud12} or regression \cite{HornsteinLopesSantosVictorLacerda2006,DeleforgeHoraud12,DeleforgeForbesHoraud13,DeleforgeForbesHoraud2014b,Luo2014}, first to learn a mapping from binaural features to the (1D or 2D) direction of a \textit{single source}, and second to infer an unknown source direction from binaural observations. These methods have the advantage that an explicit HRTF model is replaced by an implicit one, embedded in the parameters  learned during the training stage. In general, the source used for training has a wide acoustic spectrum, \textit{e.g.}, white noise (WN). A key feature common to all supervised-SSL methods is that their accuracy relies on the similarity between training and testing conditions, \textit{e.g.}, setup, room, position in the room, etc.,  rather than on the similarity between a simplified model and real world conditions.

In this paper we propose a supervised multiple-SSL method that requires neither source separation \cite{MandelWeissEllis10,DeleforgeForbesHoraud13} nor monaural segregation \cite{RomanWangBrown2003,WoodruffWang2010,WoodruffWang2012}.
Namely, we devise a regression model that directly maps a binaural spectrogram onto the direction space (azimuth and elevation) associated with a known number of simultaneously emitting sources $M$, \textit{i.e.}, \textit{co-localization}. This idea strongly contrasts with previous approaches in computational auditory scene analysis. Although strongly inspired by binaural hearing, it does not intend to mimic or emulate human perception but rather shows how new mathematical principles can be employed in practice for automated audition. The method starts with \textit{learning} the parameters of a probabilistic locally-linear regression model from associations between sound directions and binaural spectrograms. \textit{Offline learning} is followed by \textit{runtime testing}: the learnt regression parameters are used to estimate a set of unknown source directions from an observed spectrogram.

The latter is a time-series of \textit{high-dimensional} binaural feature vectors (one vector at each time frame) that 
depend on source directions, source spectra, reverberations and additive noise. While emitted spectra, reverberations, and noise strongly vary across both time and frequency, the directions are invariant, provided that the sources are static. The central idea in this paper is that the binaural features available at TF points are \textit{dominated by the source directions}, while they are \textit{perturbed} by the temporal and spectral variations of monaural cues, noise and reverberations. There are hundreds of thousands of TF points in a one second spectrogram. Source-direction information can be gathered by aggregating all these observations. 

The above formulation leads to the problem of learning a high-dimensional to low-dimensional (high-to-low) regression, which is problematic for two reasons. Firstly, the large number of parameters that needs to be estimated in this case is prohibitive \cite{Cook07,DeleforgeForbesHoraud2014}. Secondly, it is not clear how a regression, learnt with white-noise, can be used to locate natural sounds, \textit{e.g.}, speech. Common sounds yield sparse spectrograms with many TF points having no source content. 
Methods such as \cite{HornsteinLopesSantosVictorLacerda2006,talmon2011supervised,Luo2014} cannot handle natural sounds. A possible strategy could 
 be to gather binaural features from relatively long recordings, such that significant information becomes available at each frequency band. In turn, it must be assumed that there is a single source emitting over a relatively long period of time, which is unrealistic in practice. 
 
For all these reasons we propose to adopt an \textit{inverse regression} strategy \cite{Li91}. We devise a variant of the Gaussian locally-linear mapping (GLLiM) model, recently proposed \cite{DeleforgeForbesHoraud2014}. We learn a low-dimensional to high-dimensional (source-directions to binaural features) inverse regression using a training dataset composed of associations between white-noise spectrograms and sound directions. Ref. \cite{DeleforgeForbesHoraud2014} provides a closed-form expression for the \textit{forward} or \textit{direct regression}. In the context of multiple-SSL, this corresponds to the \textit{posterior distribution of source directions}, conditioned by a binaural feature vector and given the learned parameters of the inverse regression. In this paper we extend \cite{DeleforgeForbesHoraud2014} to time series of high-dimensional vectors with missing entries, \textit{i.e.}, binaural spectrograms containing TF points with no source information. We formally prove that the conditional posterior distribution that characterizes the spectrogram-to-source-directions mapping is a Gaussian mixture model whose parameters (priors, means, and covariances) are expressed analytically in terms of the parameters of the low-to-high regression. In practice we show that the proposed method robustly co-localizes sparse-spectrum sources that emit simultaneously, e.g., two speakers as illustrated in Fig.~\ref{fig:principle}-Right.


Inverse regression is also used in \cite{DeleforgeHoraud12,deleforge2014mapping} for single-SSL and in \cite{DeleforgeForbesHoraud13,DeleforgeForbesHoraud2014b} for \textit{simultaneous localization and separation}. However, in addition to learning a regression, performing both localization and separation requires a time-consuming variational EM algorithm at runtime. Moreover, in \cite{DeleforgeHoraud12,DeleforgeForbesHoraud13,DeleforgeForbesHoraud2014b} a binaural dummy head, mounted onto a pan-tilt mechanism, was used to collect datasets, \textit{i.e.}, associations between motor positions and binaural spectrograms. The emitter was kept static at a unique position in all training and test experiments, while the dummy head was rotated onto itself. The method was hence limited to theoretical conclusions rather than practical applications. In this paper we introduce a novel and elegant way of gathering data with associated ground truth\footnote{The datasets are publicly available at \url{https://team.inria.fr/perception/the-avasm-dataset/}.}.
An \textit{audio-visual source}, composed of a loud speaker and a visual marker, is manually held in front of the dummy head, and then moved from one position to the next. \textit{e.g.}, Fig.~\ref{fig:principle}. A camera is placed next to the dummy head. This setup allows to record synchronized auditory and visual signals. The horizontal and vertical positions of the loud-speaker marker in the image plane correspond to sound-source azimuth and elevation. Moreover, if a talking person is present in front of the dummy-head/camera setup, his/her mouth can be easily located using face detection methods \cite{viola2004robust,zhu2012face}. Hence, accurate ground-truth source directions are available in all cases and one can therefore quantify the performance of the proposed method. 

The remainder of the paper is organized as follows.
Section~\ref{sec:binaural} defines the concept of acoustic space and introduces the associated binaural features used in this study.
Section~\ref{sec:gllim} formulates multiple sound-source localization as a regression problem.
Section~\ref{subsec:learn} presents the model used for mapping binaural features onto multiple sound source direction vectors.
Section~\ref{subsec:mapping} extends this model to sparse binaural spectrogram inputs.
Section~\ref{sec:results} presents obtained results for single source localization and source-pair co-localization.
Section~\ref{sec:conclusion} draws conclusions and directions for future work.

%% file: spaceF.tex
\section{Binaural Features for Localization}
\label{sec:binaural}
\subsection{Acoustic Spaces}
\label{sec:space}
Let us consider a binaural system, \textit{i.e.}, two microphones plugged into the ears of a dummy head. This setup is used to record time series of \textit{binaural feature vectors} in $\mathbb{R}^D$, \textit{i.e.}, features capturing direction information associated with several sound sources. Section \ref{subsec:cues} details how such features are computed in practice. We denote by $\mathcal{D}$ a set of sound-source directions in a listener-centered coordinate frame; namely $\mathcal{D}\subset\mathbb{R}^2$ is a set of \textit{(azimuth, elevation)} angles. We denote by $\mathcal{Y}_\mathcal{D}\subset\mathbb{R}^D$ the subset of binaural feature vectors that can \textit{possibly} be captured by the microphones when a single point sound-source $m$ emits from $\xvect^{(m)}\in\mathcal{D}$. In this article we restrict the analysis to static sources. We refer to $\mathcal{Y}_\mathcal{D}$ as a \textit{simple-acoustic space} of the binaural setup \cite{DeleforgeForbesHoraud2014b}.

In this work we extend this concept to \textit{multiple} static point sound-sources that emit \textit{simultaneously} from $M$ different directions. The set of sound directions is $\mathcal{D}^M$, the $M-$th Cartesian power of $\mathcal{D}$. The \textit{multiple-acoustic space} of the binaural system is the subset $\mathcal{Y}_{\mathcal{D}^M}\subset\mathbb{R}^D$ of binaural feature vectors that can possibly be captured by the microphones when static sound sources emit from $M$ directions in $\mathcal{D}^M$. We represent an element of $\mathcal{D}^M$ by a \textit{multiple-direction} vector $\xvect\in\mathbb{R}^L$, where $L=2M$, \textit{e.g.}, $L=4$ in the case of two sources.

Notice that in general the size of binaural feature vectors is much larger than the direction set dimension, namely $D\gg L$. Hence, the acoustic space $\mathcal{Y}_{\mathcal{D}^M}$ forms an $L-$dimensional manifold embedded in $\mathbb{R}^D$.
In this article, we show how the structure of this manifold can be learned in a supervised way, and used to build an efficient multiple ($M$) sound-source localizer.

\subsection{Binaural Features}
\label{subsec:cues}
We consider a multi-direction vector $\xvect\in\mathbb{R}^L$ and we use the decomposition $\xvect=[\xvect^{(1)};\dots;\xvect^{(M)}]$, where $\xvect^{(m)}\in\mathbb{R}^2$ denotes the direction of the $m^{\text{th}}$ source and $[.;.]$ is a notation for vertical concatenation. Let:
\begin{equation}
\label{eq:source_spectrograms}
\begin{array}{l}
    \smat^{(\textrm{L})}=\{s^{(\textrm{L})}_{ft}\}_{f=1,t=1}^{F,T}\in\mathbb{C}^{F\times T} \\
     \smat^{(\textrm{R})}=\{s^{(\textrm{R})}_{ft}\}_{f=1,t=1}^{F,T}\in\mathbb{C}^{F\times T}
\end{array}
\end{equation}
be complex-valued spectrograms. These spectrograms are obtained from the left and right microphone signals using the short-time Fourier transform with $F$ frequency bands and $T$ time windows. Please see section~\ref{sec:results} for implementation details.

We consider two binaural spectrograms, namely the \textit{interaural level difference} (ILD) $\alphavect=\{\alpha_{ft}\}_{f=1,t=1}^{F,T}$,
and the \textit{interaural phase difference} (IPD),  $\phivect=\{\phi_{ft}\}_{f=1,t=1}^{F,T}$, which are defined as follows:
\begin{align}
\label{eq:alpha}
  \alpha_{ft} &= 20 \log \mid s^{(\textrm{R})}_{ft} / s^{(\textrm{L})}_{ft} \mid \in \mathbb{R}, \\
  \label{eq:phi}
  \phi_{ft} &= \exp \left( j \arg(s^{(\textrm{R})}_{ft} / s^{(\textrm{L})}_{ft}) \right) \in \mathbb{C}\equiv\mathbb{R}^2.
\end{align}
ILD and IPD cues, originally inspired by human hearing \cite{middlebrooks}, have been thoroughly studied in computational binaural sound source localization \cite{youssef2012towards}. These cues have proven their efficiency in numerous practical implementations \cite{RomanWangBrown2003,RomanWang2008,MandelWeissEllis10,LeePark2010,WoodruffWang2012,DeleforgeForbesHoraud13} as opposed to, \textit{e.g.}, the real and imaginary parts of the left-to-right spectrogram ratio, or monaural cues. Note that in our case, the phase difference is expressed in the complex space $\mathbb{C}$ (or equivalently in $\mathbb{R}^2$) to avoid problems due to phase circularity. This representation allows two nearby phase values to be close in terms of their Euclidean distance, at the cost of a redundant representation. The regression model proposed in the next sections implicitly captures dependencies between observed features through a probabilistic model, and is therefore not affected by such redundancies. This methodology was experimentally validated in~\cite{DeleforgeHoraud12}.

The \textit{binaural spectrogram} $\Ymat'=\{y'_{dt}\}_{d=1,t=1}^{D,T}$ is the concatenation of the ILD and IPD spectrograms
\begin{equation}
\Ymat'=[\alphavect;\phivect]\in\mathbb{R}^{D\times T}, 
\label{eq:bin-spect}
\end{equation}
where $D=3F$. Each frequency-time entry $y'_{dt}$ is referred to as a \textit{binaural feature}.

Let $\smat^{(m)}=\{s^{(m)}_{ft}\}_{f=1,t=1}^{F,T}$ be the spectrogram emitted by the $m^{\text{th}}$ source. The acoustic wave propagating from a source to the microphones diffracts around the dummy head. This propagation filters the signals, as expressed by the respective left- and right complex-valued HRTFs, $h^{(\textrm{L})}$ and $h^{(\textrm{R})}$ respectively. The HRTFs depend on the sound-source direction and frequency. Interestingly, HRTFs not only depend on the azimuth of the sound source but also on its elevation, due to the complex, asymmetrical shape of the head and pinna \cite{AytekinMossSimon2008}. It is shown in \cite{otani_hirahara_ise} that HRTFs mainly depend on azimuth and elevation while the distance has less impact in the far field (source distance $>1.8$ meter). The relative influence of low- and high-frequency ILD and IPD cues on direction estimation was studied in~\cite{DeleforgeForbesHoraud2014b}. By taking into account the HRTFs, the relationships between the emitted and perceived spectrograms write:
\begin{equation}
\label{eq:hrtf_model}
\begin{array}{l}
   s^{(\textrm{L})}_{ft} = \sum_{m=1}^M h^{(\textrm{L})}(f,\xvect^{(m)}) s^{(m)}_{ft} + g^{(\textrm{L})}_{ft}, \\
   s^{(\textrm{R})}_{ft} = \sum_{m=1}^M h^{(\textrm{R})}(f,\xvect^{(m)}) s^{(m)}_{ft} + g^{(\textrm{R})}_{ft}.
\end{array}
\end{equation}
Here $g^{(\textrm{L})}_{ft}$ and $g^{(\textrm{R})}_{ft}$ denote some residual noise at left- and right-microphones at $(f,t)$, which may include self noise, background noise and/or low reverberations.

Given the model (\ref{eq:hrtf_model}), if none of the sources emits at $(f,t)$, \textit{i.e.}, if $s^{(1)}_{ft}=s^{(2)}_{ft}=\dots s^{(M)}_{ft}=0$, then the corresponding binaural feature $y'_{dt}$ contains only noise, and hence it does not contain sound-source direction information. For this reason, such binaural features will be treated as \textit{missing}. Missing binaural features are very common in natural sounds, such as speech. To account for these missing features, we introduce a binary-valued matrix $\vv{\chi} = \{\chi_{dt}\}_{d=1,t=1}^{D,T}$. We use a threshold $\epsilon$ on the power spectral densities, $|s^{(\textrm{L})}_{ft}|^2$ and $|s^{(\textrm{R})}_{ft}|^2$, to estimate the entries of $\vv{\chi}$:
\begin{equation}
\chi_{dt} =
\left\{
\begin{array}{ll}
1 & \text{if} \;\; |s^{(\textrm{L})}_{ft}|^2 + |s^{(\textrm{R})}_{ft}|^2 \geq \epsilon \\
0 & \text{otherwise.} 
\end{array}
\right.
\end{equation}
The value of $\epsilon$ is estimated by averaging over time the measured noise power spectral density. 
Therefore, a binaural spectrogram:
\begin{equation}
\label{eq:Sdef}
\mathcal{S} = \{ \Ymat', \vv{\chi}\}
\end{equation}
 is fully characterized by the binaural features $\Ymat'$ and the associated activity matrix $\vv{\chi}$. 
 
We now consider the case where one or several sound sources emit at each frequency-time point $(f,t)$. The model (\ref{eq:hrtf_model}) implies that the corresponding binaural features (\ref{eq:bin-spect}) depend on the sound-source directions, but also on the emitted sounds, and microphone noises. However, while both the emitted sounds and the noise strongly vary across time and frequency, the sound-source directions are invariant, since the sources are static. With this in mind, a central postulate of this article is to consider that the binaural spectrogram entries $\{y'_{dt}\}_{d=1,t=1}^{D,T}$ are \textit{dominated} by sound-source directions and that they are \textit{perturbed} by time-frequency variations of emitted sounds and of microphone noises. In other words, these variations are viewed as \textit{observation noise}. This noise is expected to be very important, in particular for mixtures of natural sound sources. The proposed method will alleviate this issue by aggregating information over the entire binaural spectrogram $\mathcal{S}$. This typically consists of hundreds of thousands of binaural features for a one second recording of speech sources.

White-noise sources and associated binaural spectrograms are of crucial importance in our approach because the entire acoustic spectrum is being covered. In theory, white noise is a random signal with constant power spectral density over time and frequency. In practice, the recorded spectrogram of a white-noise source does not have null entries. Hence, $\vv{\chi}=\mathbf{1}_{D\times T}$ (all the entries are equal to $1$), and a white-noise binaural spectrogram does not have missing values. Let $\mathcal{S}=\{(\yvect'_1\hdots\yvect'_t\hdots\yvect'_T),\mathbf{1}_{D\times T}\}$ be a white-noise binaural spectrogram. To reduce observation noise, we define the associated \textit{binaural feature vector} as its temporal mean:
\begin{equation}
\label{eq:mean-vector}
\yvect = \frac{1}{T} \sum_{t=1}^T \yvect'_t.
\end{equation}
The set of binaural feature vectors $\yvect\in\mathcal{Y}_{\mathcal{D}^M}$ associated to sound source directions in $\xvect\in\mathcal{D}^M$ forms the multiple-acoustic space of our system. These vectors will be used to learn the relationship between input binaural signals and sound source directions.

%% file: gllimF.tex
\section{Supervised Sound Localization}
\label{sec:gllim}
In the previous section we postulated that binaural features were dominated by sound-source direction information. In this section we describe a method that allows to learn a mapping from these features to sound source directions using a regression technique. More precisely, consider a \textit{training dataset} of $N$ binaural-feature-and-source-direction pairs $\{\yvect_n,\xvect_n\}_{n=1}^{N}$, where $\yvect_n\in\mathcal{Y}_{\mathcal{D}^M}\subset\mathbb{R}^D$ is a mean binaural feature vector obtained from white noise signals (\ref{eq:mean-vector}), and $\xvect_n\in\mathcal{D}^M\subset\mathbb{R}^L$ is the corresponding multiple-direction vector, \textit{i.e.}, the azimuth and elevation of $M$ emitting sources. Notice that  we have $D\gg L$. Once the regression parameters have been estimated, it is in principle possible to infer the unknown source directions $\xvect$ from a spectrogram $\mathcal{S}=\{ \Ymat', \vv{\chi}\}$. 

However, there are two main difficulties when attempting to apply existing regression techniques to the problem of estimating sound directions from a binaural spectrogram. Firstly, the input lies in a high-dimensional space, and it is well known that high-dimensional regression is a difficult problem because of the very large number of model parameters to be estimated; this requires huge amounts of training data and may lead to ill-conditioned solvers. Secondly, many natural sounds have sparse spectrograms and hence associated binaural spectrograms often have a lot of missing entries. Nevertheless, in practice the sound localizer should not be limited to white-noise signals. Therefore, the regression function at hand, once trained, must be extendable to predict an accurate output (sound directions) from any input signal, including a sparse binaural spectrogram.

The proposed method bypasses the difficulties of high-dimensional to low-dimensional regression by considering the problem the other way around, \textit{i.e.}, low-to-high, or \textit{inverse regression}  \cite{Li91,DeleforgeForbesHoraud2014}. We assume that both the input and output are realizations of two random variables $\Yvect$ and $\Xvect$ with joint probability distribution $p(\Yvect,\Xvect;\thetavect)$, where $\thetavect$ denotes the model parameters. At training, the low-dimensional variable $\Xvect$ plays the role of the \textit{regressor}, namely $\Yvect$ \textit{is a function of} $\Xvect$ possibly corrupted by noise through $p(\Yvect|\Xvect;\thetavect)$. Hence, $\Yvect$ \textit{is assumed to lie on a low-dimensional manifold embedded in} $\mathbb{R}^D$ and parameterized by $\Xvect$. The small dimension of the regressor $\Xvect$ implies a relatively small number of parameters to be estimated, \textit{i.e.}, $\mathcal{O}[L(D+L)]$. This facilitates the task of estimating the model parameters. Once $\thetavect$ has been estimated, we show that the computation of the \textit{forward conditional density} $p(\Xvect|\Yvect;\thetavect)$ is tractable, and hence it may be used to predict the low-dimensional sound directions $\xvect$ associated with a high-dimensional mean binaural vector $\yvect$. More detailed studies on the theoretical and experimental advantages of inverse regression can be found in \cite{Li91} and \cite{DeleforgeForbesHoraud2014}.

In practice we use a method referred to as \textit{probabilistic piecewise-affine mapping} \cite{DeleforgeForbesHoraud2014b} to train a low-dimensional to high-dimensional (directions-to-binaural-features) \textit{inverse regression}. This is an instance of the more general \textit{Gaussian locally-linear mapping} (GLLiM) model \cite{DeleforgeForbesHoraud2014}, for which a Matlab implementation is publicly available.\footnote{\url{https://team.inria.fr/perception/gllim_toolbox/}.} The latter may be viewed as a generalization of mixture of experts~\cite{jordan1994hierarchical} or of joint GMM~\cite{QiaoMinematsu09}. We then derive an analytic expression for the spectrogram-to-directions \textit{forward regression}, namely the posterior distribution of sound directions conditioned by a sparse spectrogram and given the learned parameters of the inverse regression. This distribution is a Gaussian mixture model whose parameters (priors, means, and covariances) have analytic expressions in terms of the parameters of the inverse regression.

\section{Probabilistic Piecewise-Affine Mapping}
\label{subsec:learn}
This section presents the probabilistic piecewise-affine mapping model used for training. We consider inverse regression, namely \textit{from the low-dimensional space of sound directions to the high-dimensional space of white-noise spectrograms}.
Any realization
$(\yvect,\xvect)$ of $(\Yvect,\Xvect) \in \mathcal{Y} \times \mathcal{D}$ is such that $\yvect$ is
the image of $\xvect$ by one affine transformation
$\tau_k$ among $K$, plus an error term. 
 This is modeled by a missing variable $Z$ such that
$Z=k$ if and only if $\Yvect$ is the image of
$\Xvect$ by $\tau_k$. The following decomposition of the joint probability distribution is used:
\begin{align}
&p(\Yvect=\yvect,\Xvect=\xvect;\thetavect) =  \sum_{k=1}^K p(\Yvect=\yvect | \Xvect=\xvect,Z=k;\thetavect)\cdot \nonumber \\
\label{eq:joint-prob-distr}
&  p(\Xvect=\xvect | Z=k;\thetavect) \cdot p(Z=k;\thetavect).
\end{align}
The locally linear function that maps $\Xvect$ onto $\Yvect$ is
\begin{equation}
\label{eq:model_yxz}
  \Yvect=\sum_{k=1}^K \mathbb{I}(Z=k) (\Amat_k \Xvect + \bvect_k) + \Evect\:,
\end{equation}
where $\mathbb{I}$ is the indicator function and $Z$ is a hidden variable such that $\mathbb{I}(Z=k)=1$ if and only if $Z=k$, matrix
$\Amat_k\in\mathbb{R}^{D\times L}$ and vector
$\bvect_k\in\mathbb{R}^D$ are the parameters of an affine transformation $\tau_k$ and
$\Evect\in\mathbb{R}^D$ is a centered Gaussian error vector with diagonal covariance
$\Sigmamat = \diag (\sigma^2_1 \hdots \sigma^2_d \hdots \sigma^2_D )\in\mathbb{R}^{D\times D}$ capturing both the
observation noise in $\mathbb{R}^D$ and the reconstruction error
due to the local affine approximation. As already emphasized in \cite{MandelWeissEllis10}, the well known correlation between ILD and IPD cues as well as the correlation of source spectra over frequencies does not contradict the assumption that $\Sigmamat$ is diagonal, \textit{i.e.}, the Gaussian noises corrupting binaural observations are independent. This assumption was proven to be reasonable in practice, \textit{e.g.} \cite{MandelWeissEllis10}, \cite{DeleforgeForbesHoraud2014b}. Consequently we have
\begin{equation}
\label{eq:learning_model_y}
p(\Yvect=\yvect|\Xvect=\xvect,Z=k) =
 \mathcal{N}(\yvect ;\Amat_k \xvect + \bvect_k ,\Sigmamat).
\end{equation}
To make the transformations local, we associate each transformation $\tau_k$ to a region $\mathcal{R}_k\in\mathbb{R}^L$. These regions are modeled in a probabilistic way by assuming that $\Xvect$ follows
a mixture of $K$ Gaussians defined by
\begin{equation}
 p(\Xvect=\xvect |Z=k;\thetavect) = \mathcal{N}(\xvect ; \cvect_k,\Gammamat_k),
 \end{equation}
 with prior $p(Z=k;\thetavect) = \pi_k$ and with
$\cvect_k\in\mathbb{R}^L$, $\Gammamat_k\in\mathbb{R}^{L\times
L}$, and $\sum_{k=1}^K \pi_k=1$. 
This may be viewed as a compact probabilistic way of partitioning the low-dimensional space into regions. Moreover, it allows to \textit{chart} the high-dimensional space and hence to provide a piecewise affine partitioning of the data lying in this space.
To summarize, the model parameters are:
\begin{equation}
\label{eq:theta}
\thetavect = \{\{\cvect_k,\Gammamat_k,\pi_k,\Amat_k,\bvect_k\}_{k=1}^K,\Sigmamat\}.
\end{equation}
The parameter vector (\ref{eq:theta}) can be estimated via an EM algorithm using a set of associated training data $\{\yvect_n,\xvect_n\}_{n=1}^{N}$. The E-step evaluates the posteriors
\begin{equation}
\label{eq:resp-gllim}
r_{kn}^{(i)} = p(Z_{n}=k|\xvect_n,\yvect_n;\thetavect^{(i-1)})
\end{equation}
 at iteration $i$. The M-step maximizes the expected complete-data log-likelihood with respect to parameters $\thetavect$, given the observed data and the current parameters $\thetavect^{(i)}$, providing:
 \begin{equation}
 \label{eq:optimal-theta}
 \thetavect^{(i)} = \argmax_{\thetavect} \left\{
 \mathbb{E}_{(\Zvect|\Xvect,\Yvect,\thetavect^{(i-1)})}[\log p(\Xvect,\Yvect,\Zvect|\thetavect)] \right\}.
 \end{equation}
Closed-form expressions for the E- and M-steps can be found in \cite{DeleforgeForbesHoraud2014b}. We denote by $\widetilde{\thetavect}=\thetavect^{(\infty)}$ the estimated parameter vector after convergence.
The technique optimally partitions the low- and the high-dimensional spaces to minimize the reconstruction errors made by local affine transformations. It thereby captures the intrinsic structure of the acoustic space manifold $\mathcal{Y}_{\mathcal{D}^M}$.

As a further justification for learning a low-dimensional to high-dimensional regression, let us consider the number of model parameters. 
With $D=1536$ (the dimension of binaural feature vectors, see Section~\ref{sec:results}), $L=2$ (single-source localization), and $K=10$ (the number of affine transformations), there are approximately $45,000$ parameters to be estimated (\ref{eq:theta}), including the inversion of $2\times2$ full covariances $\{\mm{\Gamma}_k\}_{k=1}^{k=K}$. If instead, a high-dimensional to low-dimensional regression is learned, the number of parameters is of the order of $10^8$ and one must compute the inverse of $1536\times1536$ full covariances, which would require a huge amount of training data.

\section{From Sparse Spectrograms to Sound Directions}
\label{subsec:mapping}
We now consider the localization of natural \textit{sparse-spectrum} sounds, e.g., speech mixtures. As already mentioned, a binaural spectrogram is described by $\mathcal{S}=\{\Ymat',\chivect\}$, where $\Ymat'=\{y'_{dt}\}_{d=1,t=1}^{D,T}$ is a set of binaural features and $\chivect=\{\chi_{dt}\}_{d,t=1}^{D,T}$ is a binary-valued activity matrix. We seek the posterior density of a set of sound directions, $p(\xvect | \mathcal{S}, \widetilde{\thetavect})$, conditioned by the observed spectrogram $\mathcal{S}$ and given the estimated model parameters $\widetilde{\thetavect}$.
 We state and prove the following theorem which allows a full characterization of this density:
\begin{theorem}
\label{th:iPPAM}
Under the assumption that all the feature vectors in 
$\mathcal{S}$ are emitted from fixed directions, the following posterior distribution 
is a Gaussian mixture model in $\mathbb{R}^L$, namely
\begin{equation}
 \label{eq:spec_inv_GMM}
 p(\xvect|\mathcal{S};\widetilde{\thetavect})=\sum_{k=1}^K\nu_k\mathcal{N}(\xvect;\muvect_k,\Vmat_k).
\end{equation}
whose parameters $\{\nu_k, \muvect_k,\Vmat_k \}_{k=1}^{k=K}$ can be expressed in closed-form with respect to $\widetilde{\thetavect}$ and $\mathcal{S}$, namely:
\begin{align}
\label{eq:spectrogram_gmm_m}
\muvect_k &= \Vmat_k\biggl(\widetilde{\Gammamat}_k^{-1}\widetilde{\cvect}_k+{\sum_{d,t=1}^{D,T}}\frac{\chi_{dt}}{\widetilde{\sigma}^2_{d}}\widetilde{\avect}_{dk}(y'_{dt}-\widetilde{b}_{dk})\biggr) \:, \\
\label{eq:spectrogram_gmm_V}
\Vmat_k & = \biggl(\widetilde{\Gammamat}_k^{-1}+{\sum_{d,t=1}^{D,T}}\frac{\chi_{dt}}{\widetilde{\sigma}^2_{d}}\widetilde{\avect}_{dk}\widetilde{\avect}_{dk}\tp\biggr)^{-1}  \:,\\
\label{eq:spectrogram_gmm_rho}
\nu_k & \propto  \widetilde{\pi}_k\frac{|\Vmat_k|^{\frac{1}{2}}}{|\widetilde{\Gammamat}_k|^{\frac{1}{2}}}\exp\biggl(-\frac{1}{2}\bigl(\sum_{d,t=1}^{D,T}\displaystyle\frac{\chi_{dt}}{\widetilde{\sigma}^2_{d}}(y'_{dt}-\widetilde{b}_{dk})^2 \nonumber \\
& + \widetilde{\cvect}_k\tp\widetilde{\Gammamat}_k^{-1}\widetilde{\cvect}_k-\muvect_k\tp\Vmat_k^{-1}\muvect_k\bigr)\biggr) \:,
\end{align}
where $\widetilde{\avect}\tp_{dk}\in\mathbb{R}^L$ is the $d^{th}$ row vector of $\widetilde{\Amat}_k$, $\tilde{b}_{dk}\in\mathbb{R}$ is the  $d^{th}$ entry of $\widetilde{\bvect}_k$ and $\{\nu_k\}_{k=1}^K$ are normalized to sum to 1.
\end{theorem}
The posterior expectation of (\ref{eq:spec_inv_GMM}) can then be used to predict sound directions: 
\begin{equation}
\label{eq:post-expectation}
\widehat{\xvect}=\mathbb{E}[\xvect|\mathcal{S};\widetilde{\thetavect}]=\sum_{k=1}^K\nu_k\muvect_k \:.
\end{equation}
We refer to the resulting general sound sources localization method as \textit{supervised binaural mapping} (SBM), or SBM-$M$ where $M$ is the number of sources. Documented Matlab code for this method is available online\footnote{\url{https://team.inria.fr/perception/research/binaural-ssl/}}. 

\noindent \textbf{Proof of theorem~\ref{th:iPPAM}.}
By including the hidden variable $Z$ (section \ref{subsec:learn}) and using the sum rule, we obtain:
\begin{equation}
\label{eq:pXs_decomp}
p(\xvect|\mathcal{S};\widetilde{\thetavect}) = \sum_{k=1}^Kp(\xvect|\mathcal{S},Z=k;\widetilde{\thetavect})p(Z=k|\mathcal{S};\widetilde{\thetavect}).
\end{equation}
Since the proposed model implies an affine dependency between the Gaussian variables $\Xvect$ and $\Yvect$ given $Z$, the term $p(\xvect|\mathcal{S},Z=k;\widetilde{\thetavect})$ is a Gaussian distribution in $\xvect$. In other words, for each $k$, there is a mean $\muvect_k\in\mathbb{R}^L$ and a covariance matrix $\Vmat_k\in\mathbb{R}^{L\times L}$ such that $p(\xvect|\mathcal{S},Z=k;\widetilde{\thetavect})=\mathcal{N}(\xvect;\muvect_k,\Vmat_k)$.
Notice that $p(Z=k|\mathcal{S};\widetilde{\thetavect}) = \nu_k$ is not conditioned by $\xvect$. 

With these notations, (\ref{eq:pXs_decomp}) leads directly to (\ref{eq:spec_inv_GMM}). We now detail the computation of the GMM parameters $\{\muvect_k,\Vmat_k,\nu_k\}_{k=1}^K$.
Using Bayes inversion we have:
\begin{equation}
  \label{eq:bayes_xSZ}
  p(\xvect|\mathcal{S},Z=k;\widetilde{\thetavect})= \frac{p(\mathcal{S}|\xvect,Z=k;\widetilde{\thetavect})p(\xvect|Z=k;\widetilde{\thetavect})}{p(\mathcal{S}|Z=k;\widetilde{\thetavect})}.
\end{equation}
Since we already assumed that the measurement noise has a diagonal covariance $\Sigmamat$, the observations in $\mathcal{S}$ are conditionally independent given $Z$ and $\xvect$. 
Therefore, by omitting the denominator of (\ref{eq:bayes_xSZ}) which does not depend on $\xvect$, it follows that $p(\xvect|\mathcal{S},Z=k;\widetilde{\thetavect})$ is proportional to
\begin{align}
&p(\xvect|Z=k;\widetilde{\thetavect})  
\textstyle\prod_{d=1,t=1}^{D,T}p(y'_{dt}|\xvect,Z=k;\widetilde{\thetavect})^{\chi_{dt}}
\nonumber \\
 &=
\mathcal{N}(\xvect;\widetilde{\cvect}_k,\widetilde{\Gammamat}_k) 
\textstyle\prod_{d=1,t=1}^{D,T}\mathcal{N}(y'_{dt}|\widetilde{\avect}_{dk}\tp\xvect+\widetilde{b}_{dk},\widetilde{\sigma}^2_{d})^{\chi_{dt}}
\nonumber \\
\label{eq:decomp_xSZ}
 &= \frac{C}{|\widetilde{\Gammamat}_k|^{\frac{1}{2}}}
\exp\left(
-\frac{1}{2}\left( A + B \right) \right)\:,
\end{align}
where 
\begin{align}
\label{eq:proofA}
A &= \textstyle\sum_{d=1,t=1}^{D,T}\frac{\chi_{dt}}{\widetilde{\sigma}^2_{d}}(y'_{dt}-\widetilde{\avect}_{dk}\tp\xvect-\widetilde{b}_{dk})^2\\ 
\label{eq:proofB}
B &= (\xvect-\widetilde{\cvect}_k)\tp\widetilde{\Gammamat}_k^{-1}(\xvect-\widetilde{\cvect}_k),
\end{align}
and $C$ is a constant that depends neither on  $\xvect$ nor on $k$. Since $p(\xvect|\mathcal{S},Z=k;\widetilde{\thetavect})$ is a normal distribution in $\xvect$ with mean $\muvect_k$ and covariance $\Vmat_k$, we can write:
\begin{equation}
\label{eq:xSZ_target}
A + B = (\xvect-\muvect_k)\tp\Vmat_k^{-1}(\xvect-\muvect_k).
\end{equation}
By developing the right-hand side of (\ref{eq:xSZ_target}) and by identification with the expressions of $A$ (\ref{eq:proofA}) and $B$ (\ref{eq:proofB}), we obtain the formulae (\ref{eq:spectrogram_gmm_m}) and (\ref{eq:spectrogram_gmm_V}) for $\muvect_k$ and $\Vmat_k$ respectively.
Using Bayes inversion, one can observe that the mixture's priors $\nu_k=p(Z=k|\mathcal{S};\widetilde{\thetavect})$ are proportional to $\widetilde{\pi}_kp(\mathcal{S}|Z=k;\widetilde{\thetavect})$.
Unfortunately, we cannot directly decompose $p(\mathcal{S}|Z=k;\widetilde{\thetavect})$ into a product over $(d,t)$, as previously done with $p(\mathcal{S}|\xvect,Z=k;\widetilde{\thetavect})$. Indeed, 
while it is assumed that the frequency-time points of the observed spectrogram $\mathcal{S}$
 are independent \textit{given $\xvect$ and $Z$}, this is not true for the same observations \textit{given only $Z$}. However, we can use (\ref{eq:bayes_xSZ}) to obtain
\begin{equation}
p(\mathcal{S}|Z=k;\widetilde{\thetavect})= \frac{p(\mathcal{S}|\xvect,Z=k;\widetilde{\thetavect})p(\xvect|Z=k;\widetilde{\thetavect})}{p(\xvect|\mathcal{S},Z=k;\widetilde{\thetavect})}.
\end{equation}
The numerator is given by (\ref{eq:decomp_xSZ}) and the denominator is the normal distribution $\mathcal{N}(\xvect;\muvect_k,\Vmat_k)$. After simplifying the terms in $\xvect$, we obtain the desired expression (\ref{eq:spectrogram_gmm_rho}) for $\nu_k$. 
$\blacksquare$

%% file: resultsF.tex
\section{Experiments and Results}
\label{sec:results}
\begin{figure*}[!t]
\centering
    \begin{tabular}{c}
      \includegraphics[height = 0.35\linewidth,clip=,keepaspectratio]{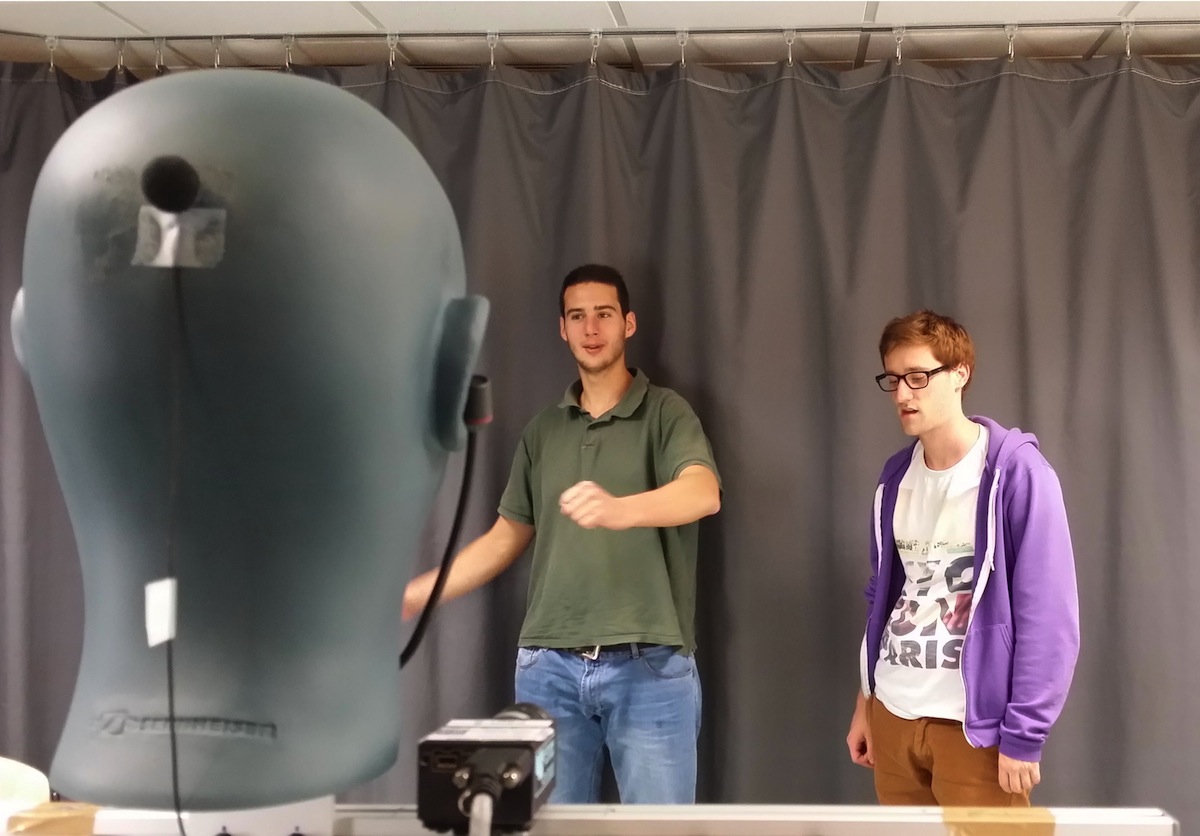}
      \includegraphics[height = 0.35\linewidth,clip=,keepaspectratio]{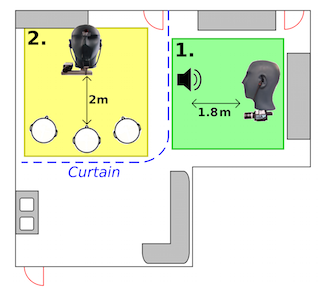}
    \end{tabular}
    \caption{\label{fig:setup} Left: A typical recording session. The two microphones plugged into the ears of the dummy head are the only ones used in this paper, although the head is equipped with four microphones. Right: Top view of the recording room. The robustness of the method with respect to room changes is validated by using different locations for training with white noise emitted by a loudspeaker (green zone \#1) and for testing with human speakers (yellow zone \#2).}
 \end{figure*}

In this section, the proposed binaural localization method is evaluated with one source ($M=1$, $L=2$) as well as with two sources ($M=2$, $L=4$). We gathered several datasets using the following experimental setup. A camera is rigidly attached next to a binaural Senheiser MKE 2002 acoustic dummy head, e.g., Fig.~\ref{fig:principle}-left and Fig.~\ref{fig:setup}-left.
We used two cameras, one with a resolution of 640$\times$480 pixels and a horizontal $\times$ vertical field of view of $28^{\circ}\times 21^{\circ}$, and another one with a resolution of 1600$\times$1200 pixels and a horizontal $\times$ vertical field of view of $62^{\circ}\times 48^{\circ}$. With these two cameras, a horizontal field of view of  $1^{\circ}$ corresponds to 23 pixels and 26 pixels, respectively. Assuming a linear relationship, pixel measurement can be conveniently converted into degrees.
The dummy-head-and-camera recording setup is placed approximately in the middle of a room whose reverberation time is $T60\approx300$~ms. Low background noise ($<28$ dBA) due to a computer fan was present as well. All the recordings (training and testing) were performed in the same room. In general, we used the same room location for training and for testing. In order to quantify the robustness of the method with respect to room locations, we carried out experiments in which we used one room location for training and another room location for testing, \textit{i.e.}, Fig.~\ref{fig:setup}-right.

The training data were obtained from a loudspeaker. The test data were obtained from a loudspeaker, and from people speaking in front of the recording device. All the training and test datasets have associated ground-truth obtained as follows. A visual pattern, which can be easily detected and precisely located in an image, was placed on the loudspeaker (Fig.~\ref{fig:principle}-middle). This setup allows us to associate a 2D pixel location with each emitted sound. Moreover, we used the Zhu-Ramanan face detection and localization method \cite{zhu2012face} that enables accurate localization of facial landmarks, such as  the mouth, in the image plane (errors made by the mouth localization method were manually corrected). Therefore, pixel positions, or equivalently sound directions, are always available with the recorded sounds. To evaluate SSL performance, we define the \textit{ground-truth-to-estimate angle} (GTEA). This corresponds to the distance between the expected sound source location (loud-speaker or mouth) and the estimated one, converted to degrees. This allows quantitative evaluation of the proposed method's accuracy, and comparison with other methods using the same datasets.

Binaural feature vectors are obtained using the short-time Fourier transform with a 64ms Hann window and 56ms overlap, yielding $T=125$ windows per second. Each time window therefore contains $1024$ samples, transformed into $F=512$ complex Fourier coefficients covering 0Hz--8kHz. We considered the following binaural feature vectors: ILD only, namely (\ref{eq:alpha}) with $D=F=512$, IPD only, namely (\ref{eq:phi}) with $D=2F= 1024$, and concatenated ILD-IPD referred to as ILPD, namely (\ref{eq:bin-spect}) with $D=3F=1536$.


Training data were gathered by manually placing the loudspeaker at $18\times24=432$ grid positions lying in the camera's field of view and in a plane which is roughly parallel with the image plane, two meters in front of the setup (Fig.~\ref{fig:principle}-middle). One-second long white-noise (WN) signals and the corresponding image positions were synchronously recorded. The training data are referred to as the \textbf{loudspeaker-WN} data. This dataset can straightforwardly be used to train a single-source localizer ($M=1$).
Importantly, training the multiple-source co-localization does 
\textit{not} require any additional data. Indeed,
the single source training dataset can also be used to generate a two-source training dataset ($M=2$),
by randomly selecting source pairs with their image-plane locations and by mixing up the two binaural
recordings.

Similarly, we gathered a test dataset by placing the loudspeaker at $9\times12=108$ positions. At each position, the loudspeaker emitted a 1 to 5 seconds random utterance from the TIMIT dataset~\cite{TIMIT}. Two-source mixtures were obtained by summing up two binaural recordings from these test data. 
As was the case with the training data, the 2D directions of the emitted sounds are available as well, thus providing the ground-truth. These test data are referred to as the \textbf{loudspeaker-TIMIT} data.
A more realistic test dataset aiming at reproducing different natural auditory scenes was gathered with one and two live speakers in front of the dummy head and camera, at a distance varying between 1.8 and 2.2 meters. We tested the following scenarios:
\begin{itemize}
\item \textbf{Moving-speaker} scenario (narrow field-of-view camera lens). A single person counts in English from 1 to 20. The person is approximatively still (small head movements are unavoidable) while she/he pronounces an isolated speech utterance, whereas she/he is allowed to wander around in between two consecutive utterances
\item \textbf{Speaking-turn} scenario (wide field-of-view camera lens). Two people take speech turns (they count from 1 to 20) with no temporal overlap, in different languages (English, Greek, Amharic). They are allowed to move between two consecutive utterances.
\item \textbf{Two-speaker} scenario (narrow field-of-view camera lens). Two people count simultaneously from 1 to 20 in different languages (English, French, Amharic) while they remain in a quasi-static position through the entire recording (see the paragraph below).
\end{itemize}

These live test data are referred to as the \textbf{person-live} dataset. All the training and test datasets, namely \textbf{loudspeaker-WN}, \textbf{loudspeaker-TIMIT}, and \textbf{person-live} are publicly available\footnote{\url{https://team.inria.fr/perception/the-avasm-dataset/}}. Notice that ground-truth 2D source directions are available with all these data, hence they can be used indifferently for training and for testing. The live recordings are particularly challenging for many reasons. The sounds emitted by a live speaker have a large variability in terms of direction of emission ($\pm30^{\circ}$), distance to the binaural dummy head ($\pm50$cm), loudness, spectrogram sparsity, etc. Moreover, the people have inherent head motions during the recordings which is likely to add perturbations. Therefore, there is an important discrepancy between the training data, carefully recorded with a loudspeaker emitting white-noise, and these test data. 

In all the \textbf{person-live} scenarios, a fixed-length analysis segment is slid along the time axis and aligned with each video frame (segments are generally overlapping). The proposed supervised binaural mapping methods, namely SBM-1 and SBM-2 (\ref{eq:post-expectation}), are applied to the segments that yield a sufficient acoustic level. The sound-source localization results obtained for each segment are then represented in their corresponding video frame.

\subsection{Single-Source Localization}
\label{subsec:single_loc}

We first evaluate our supervised binaural mapping method in the single source case ($M=1$, $L=2$), \textit{i.e.}, SBM-1. Training was done with $N=432$ binaural feature vectors associated to single source directions, using $K=32$ (13.5 points per affine transformation) and white noise recordings made with the $28^\circ\times21^\circ$ field of view camera. The overall training computation took around 5.3 seconds using Matlab and a standard PC. We compared our method with the baseline sound source localization method PHAT-histogram, here abbreviated as PHAT \cite{knapp1976generalized,Aarabi02}. PHAT-histogram estimates the \textit{time difference of arrival} (TDOA) between microphones by pooling over time generalized cross-correlation peaks, thus obtaining a pseudo probability distribution\footnote{We used the PHAT-histogram implementation of Michael Mandel, available at http://blog.mr-pc.org/2011/09/14/messl-code-online/. This TDOA estimator has a sub-sample accuracy, allowing for non-integer sample delays.}. In all experiments, the same sampling frequency (16kHz) and the same sound length is used with PHAT and with our method. A linear regression was trained to map TDOA values obtained with PHAT, onto the horizontal image axis using \textbf{loudspeaker-WN} training data\footnote{A linear dependency between TDOA values obtained with PHAT using a single white-noise source and its horizontal pixel position was observed in practice.}. Notice that PHAT, as well as all binaural TDOA-based localization methods, cannot estimate the vertical position/direction. The few existing 2D sound source localization methods in the literature, e.g., \cite{kulaib20092d}, could not be used for comparison (\cite{kulaib20092d} relies on artificial ears with a spiral shape).

\subsubsection{Loudspeaker Data}
The single-source localization results using the \textbf{loudspeaker-TIMIT} dataset are summarized in Table-\ref{tab:results_1source}. The best results are obtained using the proposed method SBM-1 and ILPD spectrograms, \textit{i.e.}, (\ref{eq:bin-spect}). The largest GTEA is then $3.94^\circ$ which corresponds to $90$ pixels. The largest GTEA with PHAT is $9^{\circ}$, and PHAT yields 14 GTEA values (out of 108 tests) which are larger than $5^\circ$. 
The proposed method outperforms PHAT, both in terms of the average error of the inliers and in the percentage of outliers, while localizing the source in 2D.

The average CPU time of our method implemented in MATLAB, using ILPD spectrograms, is of 0.23s for one second utterances, which is comparable to PHAT's CPU time. We measured the effect of introducing the binary activity matrix $\vv{\chi}$ (see Section \ref{subsec:cues} and eq. (\ref{eq:spec_inv_GMM})-\ref{eq:spectrogram_gmm_rho})). For the single-source case, taking the entire spectrogram into account instead increased the localization errors of our method by $7\%$ in average.

\begin{table}[h!]
\centering
   \caption {\label{tab:results_1source} Single-source localization results with the \textbf{loudspeaker-TIMIT} test data, using the proposed method (with ILD-, IPD-, and ILPD-spectrograms) and PHAT. The average and standard deviation (Avg$\pm$Std) of the GTEA are estimated over 108 localizations. The fourth column provides the percentage of GTEA greater than $5^\circ$. 
   }
      \footnotesize
   \begin{tabular}{|c|c|c|r|c|}
      \hline
      Method 	& Azimuth ($^\circ$)   & Elevation  ($^\circ$)   & $>5^\circ$ (\%) & CPU time (s) \\
      \hline
      ILPD   	& 0.96$\pm$0.73  & 1.07$\pm$0.92  & 0.0  & 0.23\\
      ILD       & 1.20$\pm$0.99  & 1.09$\pm$1.45  & 1.9  & 0.08\\
      IPD     	& 1.05$\pm$0.90  & 1.46$\pm$1.70  & 5.6  & 0.15\\
      PHAT      & 2.80$\pm$2.25  & -              & 14   & 0.37\\
      \hline
   \end{tabular}
\end{table}

\begin{figure}[b!]
  \vspace{-4mm}
$\begin{array}{cc}
  \includegraphics[height= 0.79\linewidth,clip=,keepaspectratio]{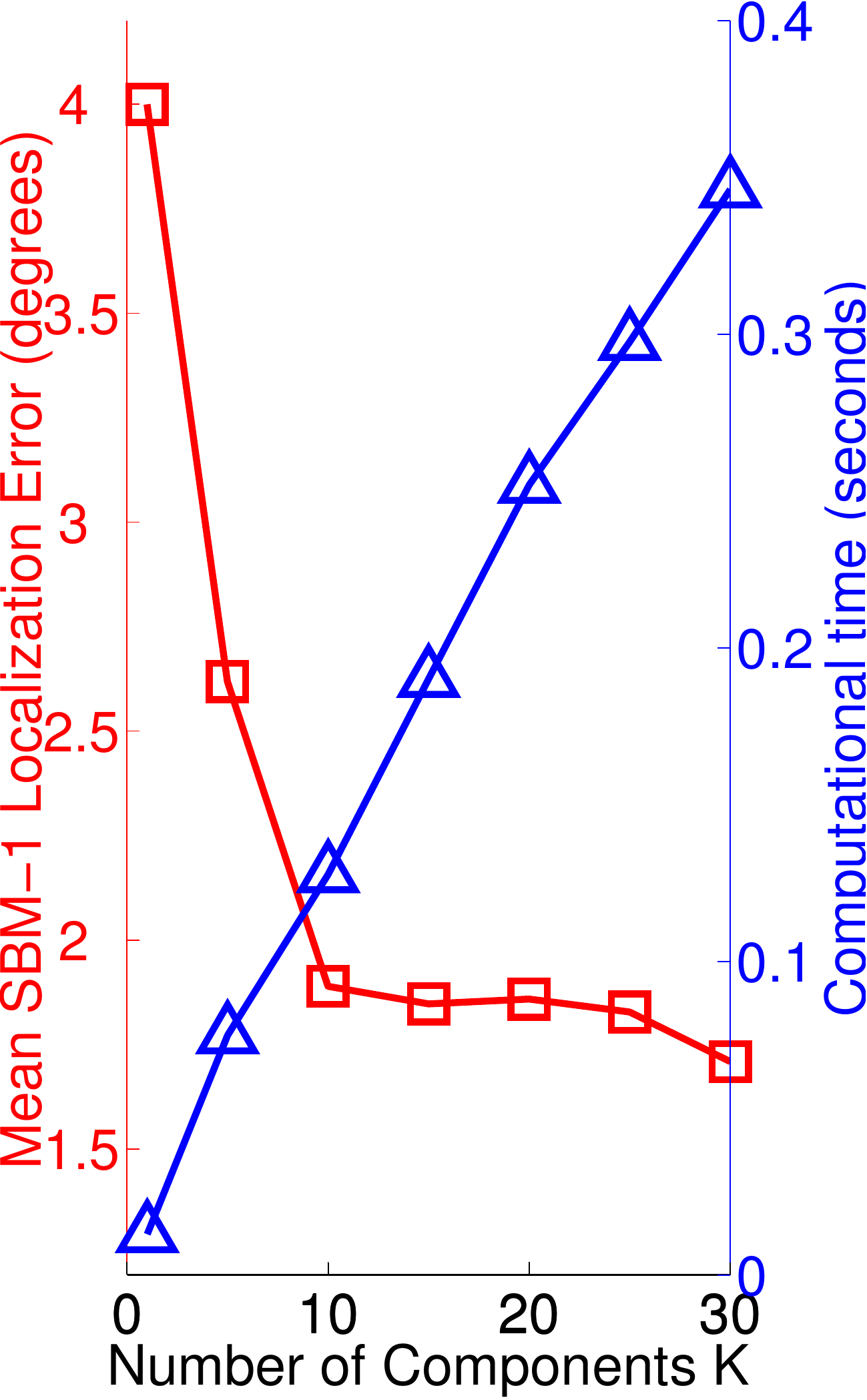} &
    \includegraphics[height = 0.79\linewidth,clip=,keepaspectratio]{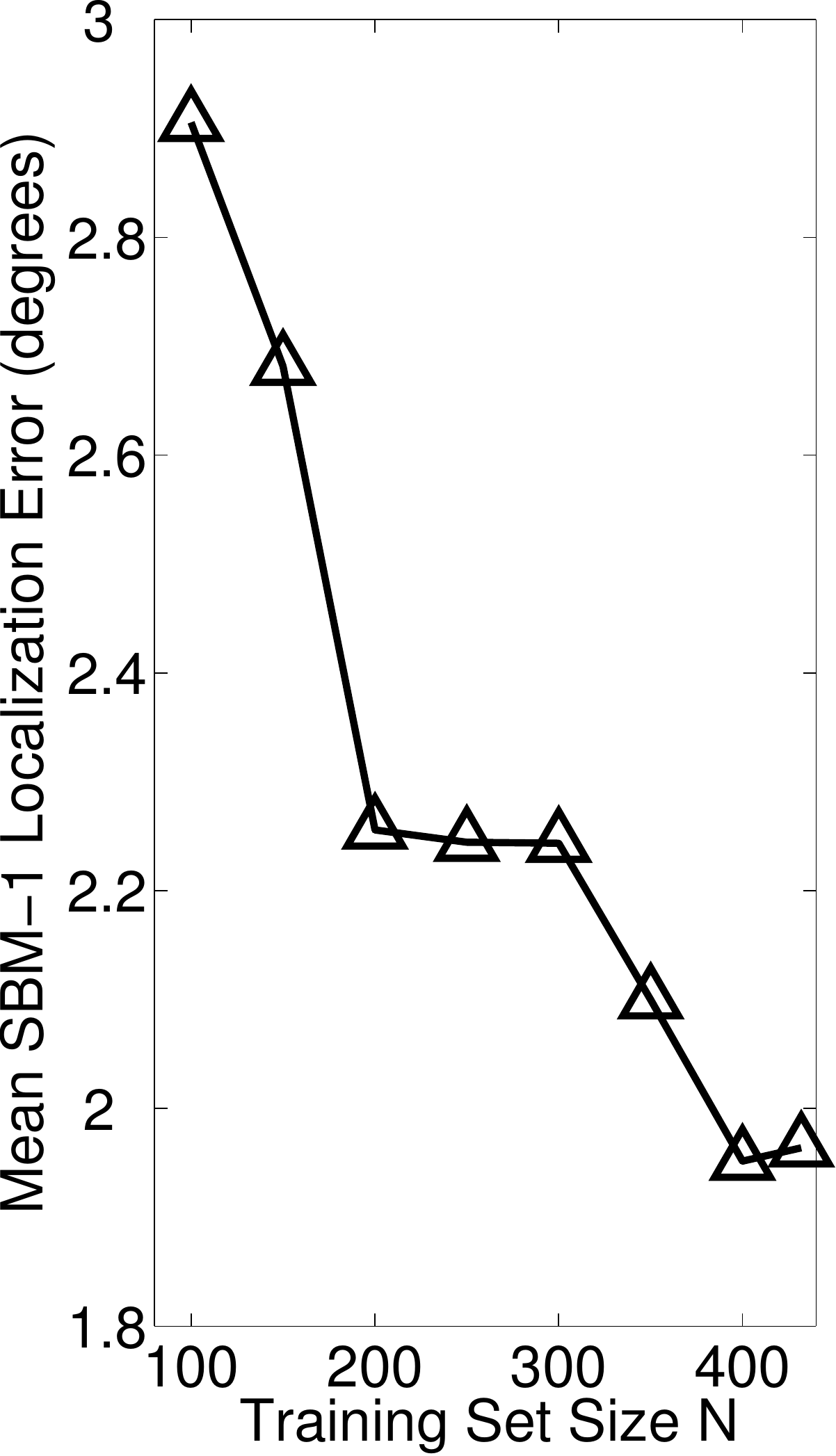} \\
\end{array}$
  \caption{\label{fig:error_plots_M1} Left: Influence of $K$ on the mean GTEA and localization time of a 1 second speech source using the proposed supervised binaural mapping method ($M=1$,$N=432$). Right: Influence of $N$ on the mean GTEA of a 1 second speech source ($K=32$).}
  \vspace{-4mm}
\end{figure}

Fig.~\ref{fig:error_plots_M1} shows the influence of the free parameters of training on the proposed method, namely the number of Gaussian component $K$ and the size of the training set $N$. As can be seen in Fig.~\ref{fig:error_plots_M1}-left, $K$ can be chosen based on a compromise between computational time and localization accuracy. Note that in this example, results do not improve much for initial $K$ values larger than $10$. This is notably because too high values of $K$ lead to degenerate covariance matrices in classes where there are too few samples. Such classes are simply removed along the execution of the algorithms, thus reducing the final value of $K$. As can be seen in Fig.~\ref{fig:error_plots_M1}-right, the number of training points $N$ has a notable influence on the localization accuracy. This is because the larger the $N$, the larger the angular resolution of the training set. However, using $N=100$ instead of $N=432$  increases the average GTEA by less than $1^{\circ}$. This suggests that a less dense grid of points could be used for simple, practical training.  While manually recording 432 positions took 22 minutes, a training set of 100 positions can be recorded in 5 minutes.

\begin{figure*}[h!t!b!]
\centering
      \begin{tabular}{ccccc}
	\includegraphics[width = 0.18\linewidth,clip=,keepaspectratio]{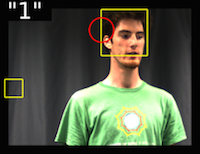}  &
	\includegraphics[width = 0.18\linewidth,clip=,keepaspectratio]{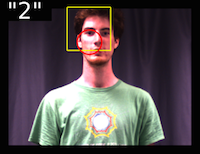}  &
	\includegraphics[width = 0.18\linewidth,clip=,keepaspectratio]{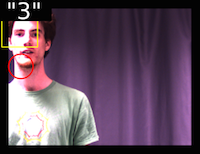}  &
	\includegraphics[width = 0.18\linewidth,clip=,keepaspectratio]{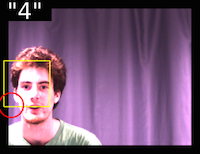}  &
	\includegraphics[width = 0.18\linewidth,clip=,keepaspectratio]{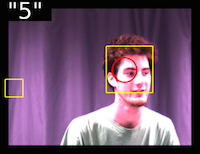}  \\
	\includegraphics[width = 0.18\linewidth,clip=,keepaspectratio]{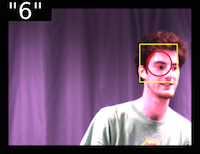}  &
	\includegraphics[width = 0.18\linewidth,clip=,keepaspectratio]{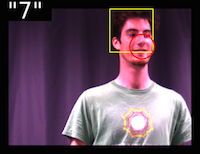}  &
	\includegraphics[width = 0.18\linewidth,clip=,keepaspectratio]{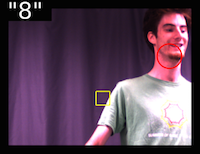}  &
	\includegraphics[width = 0.18\linewidth,clip=,keepaspectratio]{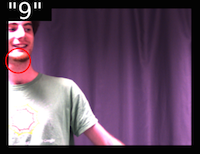}  &
	\includegraphics[width = 0.18\linewidth,clip=,keepaspectratio]{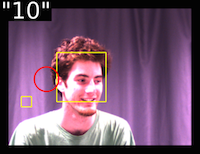} \\
	\includegraphics[width = 0.18\linewidth,clip=,keepaspectratio]{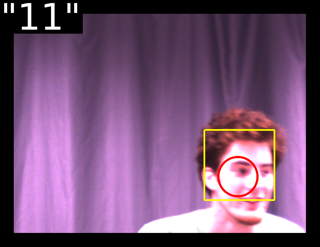} &
	\includegraphics[width = 0.18\linewidth,clip=,keepaspectratio]{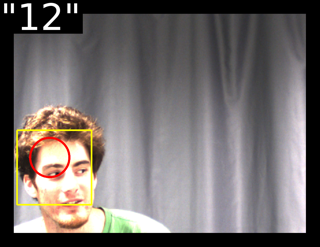} &
	\includegraphics[width = 0.18\linewidth,clip=,keepaspectratio]{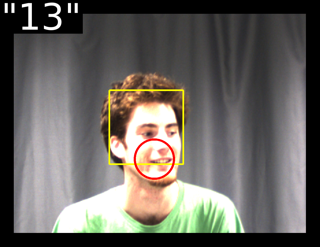} &
	\includegraphics[width = 0.18\linewidth,clip=,keepaspectratio]{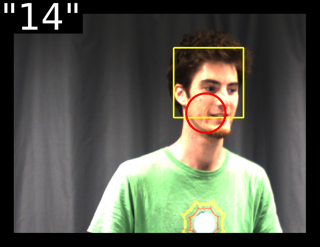} &
	\includegraphics[width = 0.18\linewidth,clip=,keepaspectratio]{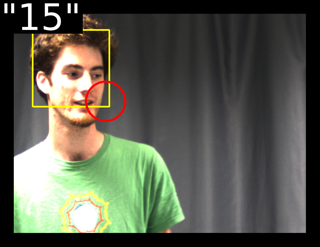} \\
	\includegraphics[width = 0.18\linewidth,clip=,keepaspectratio]{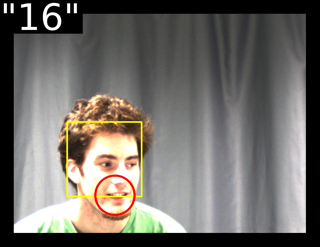} &
	\includegraphics[width = 0.18\linewidth,clip=,keepaspectratio]{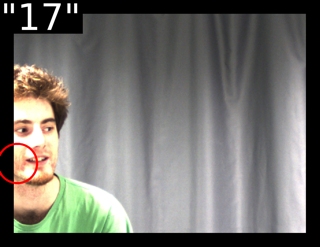} &
	\includegraphics[width = 0.18\linewidth,clip=,keepaspectratio]{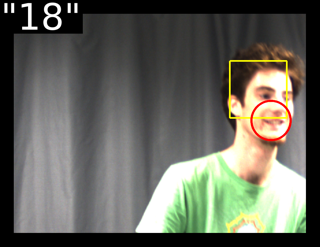} &
	\includegraphics[width = 0.18\linewidth,clip=,keepaspectratio]{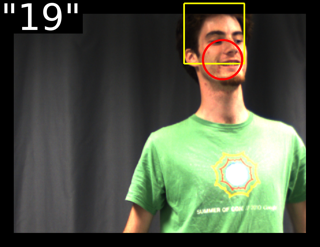} &
	\includegraphics[width = 0.18\linewidth,clip=,keepaspectratio]{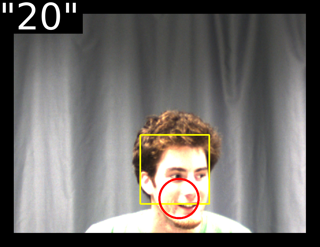} \\
      \end{tabular}
      \caption{\label{fig:counting1_scenario} The moving-speaker scenario. The person is static while he pronounces a number (written in white) and he moves between two numbers. The red circles show the position found with our method. The yellow squares correspond to faces detected with \cite{viola2004robust}. The full video is available at https://team.inria.fr/perception/research/binaural-ssl/.}
\end{figure*}

\subsubsection{Person-Live Data}
To illustrate the effectiveness of the proposed framework in real world conditions, we applied the SBM-1 method to the \textbf{moving-speaker} scenario of the \textit{person-live} dataset. A 720~ms sliding segment was used, allowing to obtain a sound source direction for each video frame with a sufficient acoustic level.  Fig.~\ref{fig:counting1_scenario} shows an example frame for each pronounced number. Note that in this experiment as well as in all the person-live experiments, the example frames are manually selected so that the corresponding analysis segments are roughly centered on the uttered numbers. Segments containing only a small part of a utterance, two consecutive utterances, or a high amount of late reverberations generally yielded unpredictable localization results. This can be observed in the online videos\footnote{\url{https://team.inria.fr/perception/research/binaural-ssl/}}. This could be addressed by using a more advanced speech activity detector to adjust the size and position of the analysis segment, as well as a tracker that takes into account past information.

The estimated location of the sound-source is shown with a red circle. For comparison, Fig.~\ref{fig:counting1_scenario} also shows face localization estimates obtained with an efficient implementation of the Viola-Jones face detector \cite{viola2004robust}. This implementation has CPU performance comparable to our method while the more precise Zhu-Ramanan face detector \cite{zhu2012face} used for ground-truth annotation is two orders of magnitude slower. Our method localizes the 20 uttered numbers, with an average GTEA and standard deviation of $1.8^\circ\pm1.6^\circ$ in azimuth and $1.6^\circ\pm1.4^\circ$ in elevation. The largest GTEA (number ``10"), is of $6.6^{\circ}$ in azimuth and $3.4^{\circ}$ in elevation. For comparison, the average azimuth localization error with PHAT-histogram is $2.7^{\circ}\pm1.9^\circ$ on the same data, with a maximum error of $6.6^\circ$. Interestingly, the Viola-Jones method \cite{viola2004robust} correctly detects and localizes the face in 16 out of 20 tests, but it fails to detect the face for ``8", ``9", ``17" and ``18". This is because the face is only partially visible in the camera field of view, or has changing gaze directions. Moreover, it features several false detections (``1", ``5", ``8", ``10").
These examples clearly show that our method may well be viewed as complementary to visual face detection and localization: it localizes a speaking face even when the latter is only partially visible and it discriminates between speaking and non-speaking faces.

\begin{figure*}
\centering
      $\begin{array}{ccccc}
	\includegraphics[width = 0.18\linewidth,clip=,keepaspectratio]{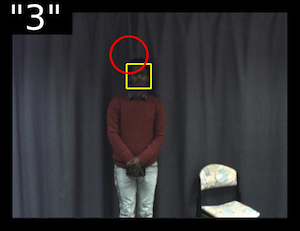}  &
	\includegraphics[width = 0.18\linewidth,clip=,keepaspectratio]{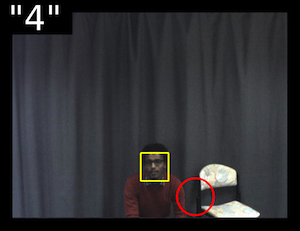}  &
	\includegraphics[width = 0.18\linewidth,clip=,keepaspectratio]{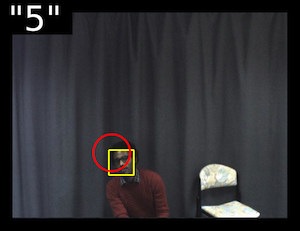}  &
	\includegraphics[width = 0.18\linewidth,clip=,keepaspectratio]{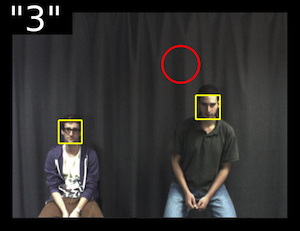}  &
	\includegraphics[width = 0.18\linewidth,clip=,keepaspectratio]{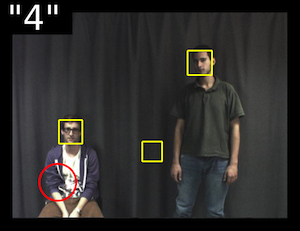}  \\
	\includegraphics[width = 0.18\linewidth,clip=,keepaspectratio]{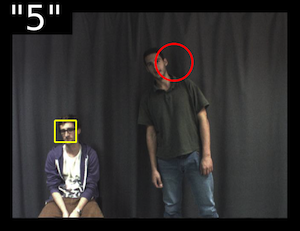}  &
	\includegraphics[width = 0.18\linewidth,clip=,keepaspectratio]{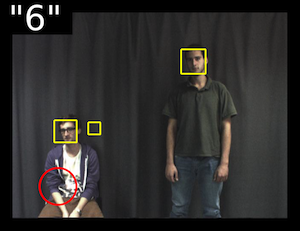}  &
	\includegraphics[width = 0.18\linewidth,clip=,keepaspectratio]{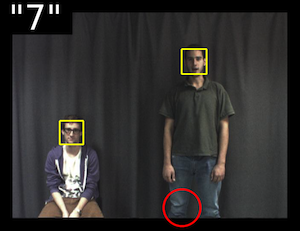}  &
	\includegraphics[width = 0.18\linewidth,clip=,keepaspectratio]{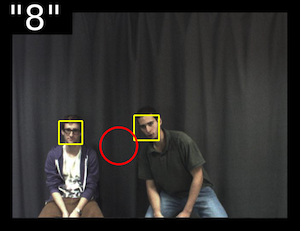}  &
	\includegraphics[width = 0.18\linewidth,clip=,keepaspectratio]{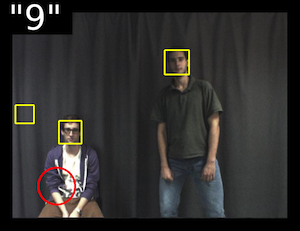}
      \end{array}$
      \caption{\label{fig:counting1_robust} Examples of localization results with a wider field-of-view camera ($62^\circ\times 48^\circ$) in a location that is different than the location used for training. Notice that overall, the method is relatively robust to changes in the room impulse response. The red circles show the position found with our method. The yellow squares show the results of the Viola-Jones face detector\cite{viola2004robust}. While the accuracy in azimuth is not significantly affected, the accuracy in elevation is significantly degraded. The full video is available at https://team.inria.fr/perception/research/binaural-ssl/.}
\end{figure*}

\subsubsection{Robustness to Locations in the Room}
\label{subsubsec:robust}

The proposed method trains a binaural system in a reverberant room and at a specific room location. In such an echoic environment, the learned parameters are therefore likely to capture the HRTF as well as the room impulse response. We remind that the method essentially relies on the similarity between training and testing conditions, rather than on the similarity between a simplified acoustic model and real world conditions.
In previous experiments, the training and testing positions were almost the same.
The objective of this experiment is to verify
whether the proposed method yields some degree of robustness to changes in room impulse responses, \textit{e.g.}, the training and testing occur at two different positions in the room. Moreover, for these experiments we used a camera with a larger field of view, namely $62^\circ\times48^\circ$. Figure \ref{fig:setup}-right shows a top view of the recording room with a training zone and a test zone. The microphone-to-emitter distance vary from 2~m (training) to approximately 1.8~m (testing). The SBM-1 method is applied to the \textbf{speaking-turn} scenario for testing. The procedure already described above is used to train the model, to localize sounds online and to select example frames. This time, a 1000~ms analysis segment is used, as it improves the overall performance.

Figure \ref{fig:counting1_robust} shows some of the results obtained with human speakers using single-source training. Out of 23 uttered numbers, the average azimuth error is $4.7^\circ\pm2.7^\circ$ with a maximum error of $9.9^\circ$. The average elevation error is $7.3^\circ\pm4.6^\circ$ excluding three outliers having an error larger than $15^\circ$ degrees, \textit{e.g.}, Fig. \ref{fig:counting1_robust}, second row, third column. Note that the increased camera field of view decreased the angular resolution of the training set by a factor $2.2$ (1 point every $2.5\circ$ in azimuth and elevation). This decreased resolution yields a slight increase in azimuth localization error, consistently with observations in Fig.~\ref{fig:error_plots_M1}-right. But overall, the azimuth accuracy does not seem to be significantly affected by changes of microphone locations in the room. On the other hand, the elevation accuracy is significantly decreased, with errors 4.6 times larger, and $8.7\%$ of outliers instead of none. This suggests that making elevation estimation more robust would require combining training data from different real and/or simulated rooms.

For comparison, the baseline algorithm PHAT is used on the same test data. As in previous experiments, a linear dependency between TDOA values estimated by PHAT and the horizontal image axis was observed. For fairness, this dependency was modeled by learning a linear regression model using the white-noise recordings from the training zone (Figure \ref{fig:setup}-right). The average azimuth error of PHAT over 23 uttered numbers is $5.4^\circ\pm2.9^\circ$ with a maximum error of $11.8^\circ$. In this realistic scenario with different training and testing locations, the proposed approach still performs better and more robustly than the baseline PHAT-histogram method in azimuth only, while it estimates the elevation as well.

\subsection{Two-Source Localization}
\label{subsec:pair_loc}
In this section, we present a key result of the proposed framework: it successfully maps binaural features to the direction-pair of two simultaneous sound sources \textit{without} relying on sound source separation. This is solely achieved based on supervised learning. This instance corresponds to the case $M=2$, $L=4$, and is thus referred to as \textit{SBM-2}.
Training was based on $N=20,000$ binaural feature vectors associated to source-pair directions, using $K=100$ (200 points per affine transformation). The overall training took 54 minutes using Matlab and a standard PC.

We compared SBM-2 to three other multiple SSL methods: PHAT-histogram \cite{Aarabi02}, MESSL \cite{MandelWeissEllis10} and VESSL \cite{DeleforgeForbesHoraud13}. PHAT-histogram can be used to localize more than one source by selecting $M$ peaks in the histograms. MESSL is an EM-based algorithm which iteratively estimates a binary mask and a TDOA for each source. The version of MESSL used here is initialized by PHAT-histogram and includes a garbage component as well as ILD priors to better account for reverberations. As in previous section, a linear regressor was trained to map PHAT's and MESSL's TDOA estimates to horizontal pixel coordinates.
The supervised method VESSL may be viewed as an extension of SBM-1 to multiple sound source separation and localization. Similarly to MESSL, VESSL uses a variational EM procedure to iteratively estimate a binary mask and a 2D direction for each source. It was trained using the single-source loudspeaker-WN dataset ($N'=432$ ILPD feature vectors) and $K'=32$ affine component. This method, as well as PHAT-histogram and MESSL, strongly rely on the assumption that emitting sources are sparse, so that a single source dominates each time-frequency bin (WDO). This is not the case of SBM-2, since it is trained with strongly overlapping white-noise mixtures.

\begin{table*}
\centering
   \caption {\label{tab:results} {Source pair localization results for different mixture types using SBM-2  and different methods. Error averages and standard deviations (Avg$\pm$Std) are showed in degrees. Avgs and Stds are only calculated over inlying estimates. Estimates are considered outliers if their distance to ground-truth is more than $15^\circ$. Percentages of outliers are given in columns ``out''. }}
   \vspace{-2mm}
   \begin{tabular}{|cc|ccc|ccc|ccc|ccc|}
      \hline
      \multicolumn{2}{|c}{Mixture} & \multicolumn{3}{c}{WN+WN} & \multicolumn{3}{c}{WN+S (WN)} & \multicolumn{3}{c}{WN+S (S)} &  \multicolumn{3}{c|}{S+S}\\
      \hline
      \multicolumn{2}{|c|}{Method}    & azimuth & elevation & out & azimuth & elevation & out & azimuth & elevation& out & azimuth & elevation & out \\
      \cline{3-14}
      \multicolumn{2}{|c|}{SBM-2.ILPD}                       & {\bf0.76$\pm$0.84} & {\bf0.99$\pm$1.11} & {\bf 0.0} & {\bf 0.83$\pm$1.18} & {\bf 0.69$\pm$1.00} & {\bf 0.0}
                                                               & {\bf3.22$\pm$3.11} & {\bf3.60$\pm$3.21} & {\bf 9.1} & {\bf 1.39$\pm$1.40} & {\bf 1.99$\pm$2.30} & {\bf 0.4} \\
      \multicolumn{2}{|c|}{SBM-2.ILD}                        & 1.03$\pm$1.51 & 1.08$\pm$1.55 & 0.1 & 1.19$\pm$1.89 & 1.15$\pm$1.74 & 0.7
                                                               & 3.28$\pm$3.03 & 3.74$\pm$3.31 & 7.8 & 2.19$\pm$2.69 & 2.48$\pm$2.85 & 3.1 \\
      \multicolumn{2}{|c|}{SBM-2.IPD}                        & 1.14$\pm$1.28 & 1.47$\pm$1.94 & 0.0 & 1.01$\pm$1.28 & 0.88$\pm$1.33 & 0.3
                                                               & 3.71$\pm$3.32 & 4.09$\pm$3.36 & 9.1 & 2.00$\pm$2.08 & 2.58$\pm$2.73 & 1.3 \\
      \multicolumn{2}{|c|}{VESSL\cite{DeleforgeForbesHoraud13}}          & 3.20$\pm$3.47 & 3.51$\pm$3.65 & 17  & 0.62$\pm$1.13 & 0.73$\pm$1.10 & 1.6
                                                                         & 5.90$\pm$3.91 & 5.35$\pm$3.84 & 35  & 3.47$\pm$3.41 & 3.69$\pm$3.57 & 11  \\
      \multicolumn{2}{|c|}{MESSL\cite{MandelWeissEllis10}}               & 4.11$\pm$3.88 & $-$ & 24  & 2.85$\pm$3.99 & $-$ & 25
                                                                         & 6.66$\pm$4.26 & $-$ & 28  & 4.05$\pm$3.90 & $-$ & 19 \\
      \multicolumn{2}{|c|}{PHAT\cite{Aarabi02}}                          & 4.01$\pm$3.89 & $-$ & 24  & 2.86$\pm$3.98 & $-$ & 25
                                                                         & 6.53$\pm$4.26 & $-$ & 28  & 4.09$\pm$3.85 & $-$ & 18 \\
      \hline
   \end{tabular}
   \vspace{-4mm}
\end{table*}

\subsubsection{Loudspeaker Data}
The methods were first tested using the loudspeaker datasets. We tested $1000$ source-pair mixtures of the following types: white-noise + white-noise (WN+WN), white-noise + speech (WN+S) and speech + speech (S+S). Each mixture was cut to last 1 second. The average amplitude ratio of source-pairs was $0\pm0.5$dB in all mixtures. The maximum azimuth and elevation distances between two test sources was $20^\circ$, and the minimal distance was $1.5^\circ$.

Table~\ref{tab:results} displays errors in azimuthal/horizontal and  elevation/vertical localization, in degrees. For WN+S mixtures, localization errors for white-noise sources (WN) and speech (S) sources are shown separately. Generally, the SBM-2 outperforms PHAT, MESSL, and VESSL in terms of accuracy, while localizing sources in 2D. Again, best results were obtained using ILPD features. SBM-2 performs best in WN+WN mixtures. This is expected because it corresponds to mixtures used for training the algorithm. However, the proposed method also yields good results in speech localization, even in the very challenging WN+S mixtures, despite an average speech-to-noise ratio of $0\pm0.5$dB. It also yields good results for S+S mixtures, even though in this case both sources are sparse. This shows that aggregating a large number of binaural features in the time-frequency plane is a successful strategy to overcome the high variability of emitted signals which affects binaural features. Moreover, introducing the binary activity matrix $\vv{\chi}$ reduced the average localization errors of our method by $25\%$ for white-noise sources and $15\%$ for speech sources. Although both SBM-2 and VESSL are based on supervised learning, our method yields significantly better results than the VESSL. This demonstrates the prominent advantage of relaxing the WDO assumption for multiple sound-source localization. The proposed method reduces the localization error by $60\%$ with respect to the second best method VESSL in a two-speaker scenario. Such a gain can be critical to correctly identify speaking people in a dense cocktail party scenario, \textit{e.g}, two people talking one meter from each other, 2 meters away from the setup.

The fact that VESSL, MESSL and PHAT perform poorly in WN+WN mixtures is expected, because then the WDO assumption is strongly violated. In WN+S mixtures, they show better performance in localizing the white noise sound source than the speech source. This can be explained by the sparsity of the speech signal. This implies that most binaural cues in the time-frequency plane are generated by the white noise source only. These cues are correctly clustered together assuming WDO, and can then be accurately mapped to the correct source direction. In the particular case of WN localization in WN+S mixtures, the average azimuth error of VESSL is even slightly lower than that of SBM-2. Possibly, this is because VESSL uses 32 affine components and a single source training ($L=2$), while SBM-2 uses 100 affine components and a two sources training ($L=4$). The average angular area per source covered by affine transformations is thus higher in VESSL ($4.3^\circ\times4.3^\circ$ on an average) than in SBM-2 ($7.7^\circ\times7.7^\circ$ on an average).

Computational times of PHAT, MESSL, VESSL ($K'=32$) and SBM-2.ILPD ($K=100$) for a one second test mixture were respectively $0.27\pm0.01s$, $10.4\pm0.1$s, $46.7\pm1.2$s and $2.2\pm0.1s$ using MATLAB and a standard PC. With proper optimization, SBM-2 is therefore suitable for real-time applications. This is not the case for MESSL and VESSL, due to their iterative nature. While the offline training of SBM methods requires a computationally costly EM procedure, the localization is very fast using the closed-form expression (\ref{eq:post-expectation}).

\begin{figure}[b!]
  \vspace{-4mm}
$\begin{array}{cc}
  \includegraphics[height= 0.79\linewidth,clip=,keepaspectratio]{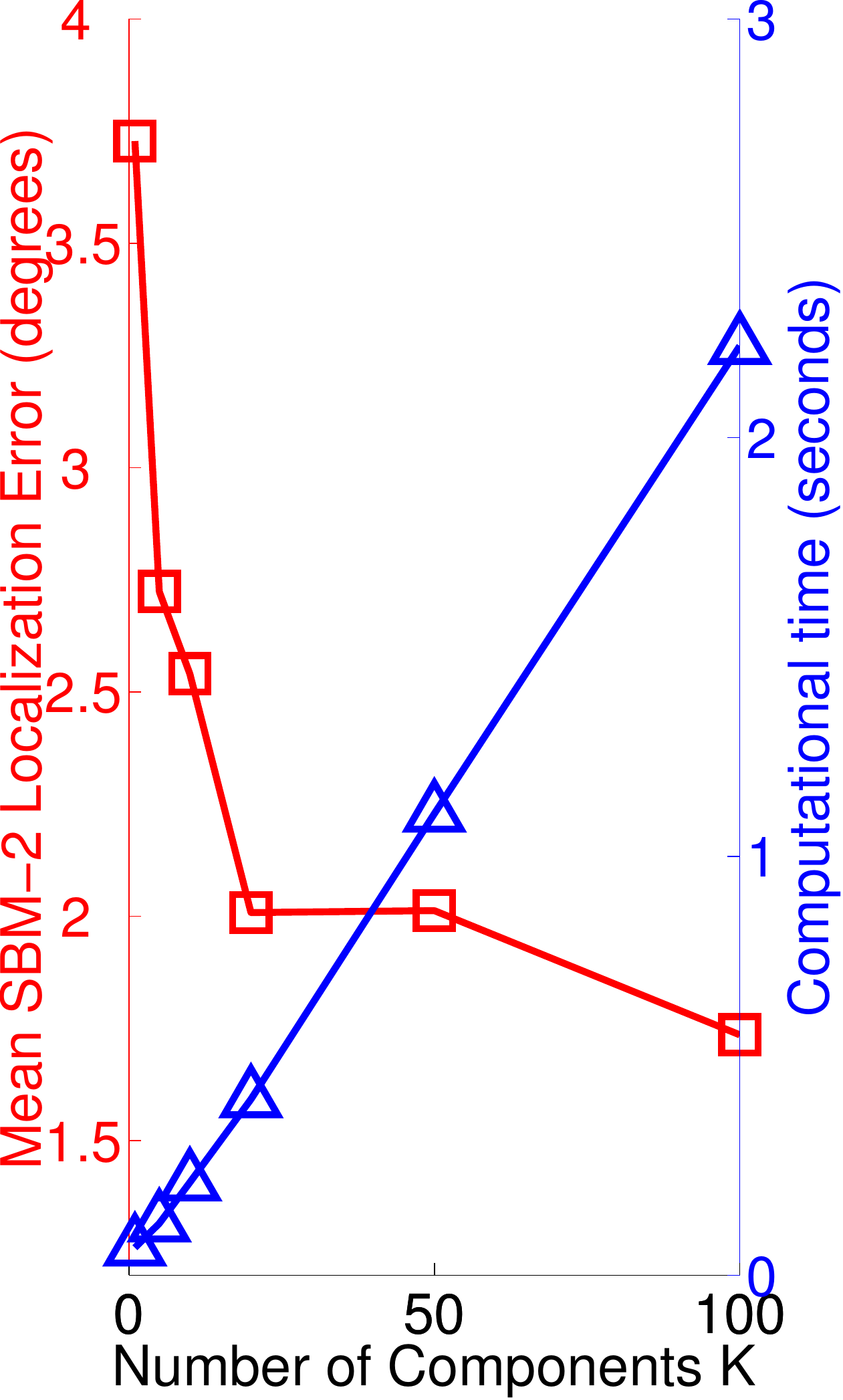} &
    \includegraphics[height = 0.79\linewidth,clip=,keepaspectratio]{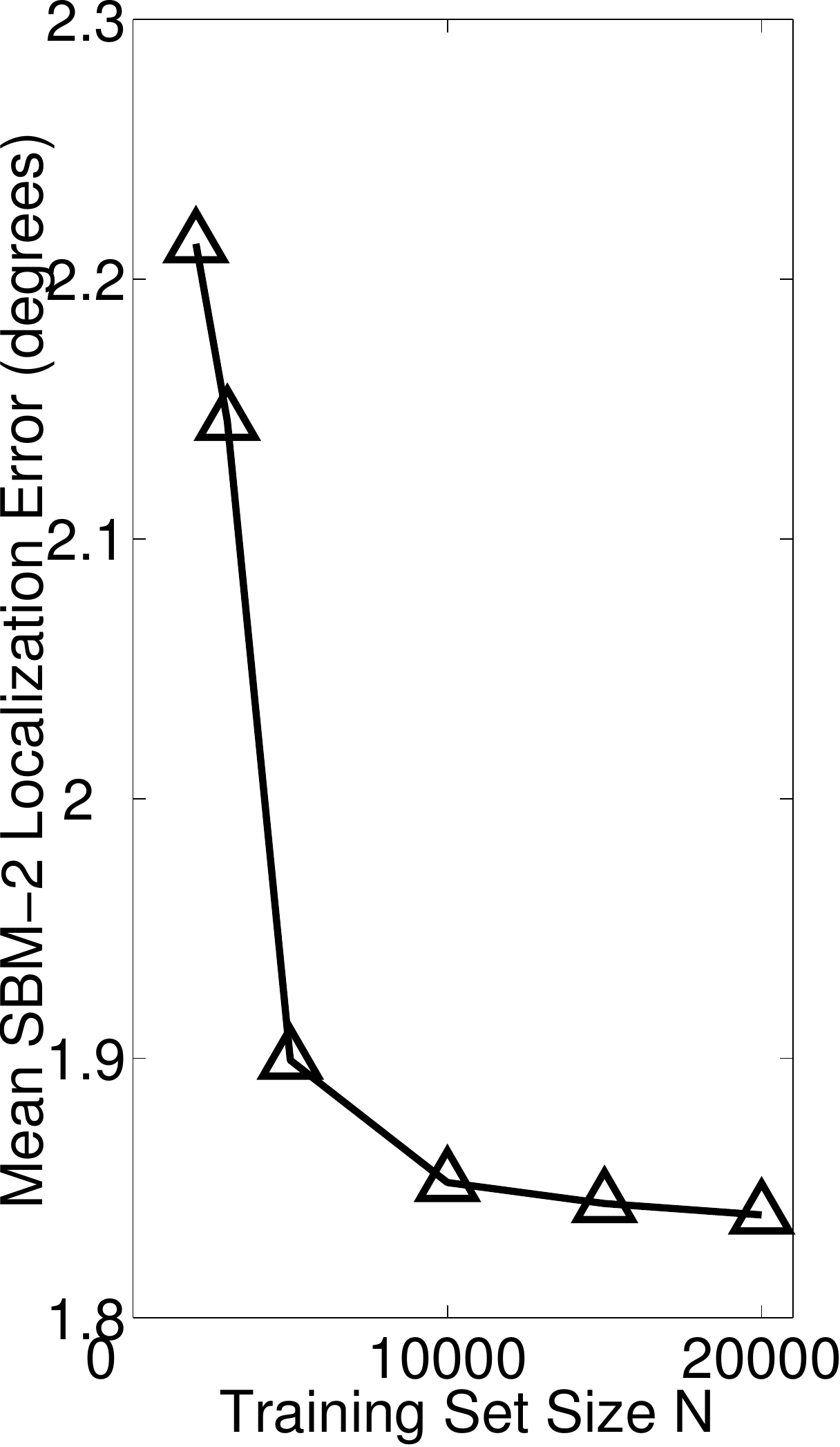} \\
\end{array}$
  \caption{\label{fig:error_plots_M2} Left: Influence of $K$ on the mean GTEA and localization time of a 1 second S+S mixture using the proposed supervised binaural mapping method ($M=2,N=20,000$). Right: Influence of $N$ on the mean GTEA of a 1 second S+S mixture ($K=100$).}
  \vspace{-4mm}
\end{figure}

As in previous section, we tested the influence of the number of affine components $K$ and training set size $N$ on SBM-2 performance. By Fig.~\ref{fig:error_plots_M2}-left, $K$ can again be tuned based on a trade-off between computation time and accuracy. Choosing $K=20$ brings down the co-localization time of a 1 second mixture to 0.42 seconds, while increasing the localization error by only $6.5\%$ relative to $K=100$. Fig.~\ref{fig:error_plots_M2}-right shows that localization error increases when $N$ decreases. However, using $N=5,000$ increases the mean localization error by only $3.2\%$ relative to $N=20,000$. Again, this suggests that a less dense grid of points can be used for faster training (a training set of 100 positions can be recorded in 5 minutes and allows $N=5,050$ source pairs).

\begin{figure*}
\centering
      $\begin{array}{ccccc}
	\includegraphics[width = 0.18\linewidth,clip=,keepaspectratio]{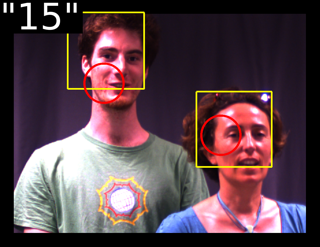}  &
	\includegraphics[width = 0.18\linewidth,clip=,keepaspectratio]{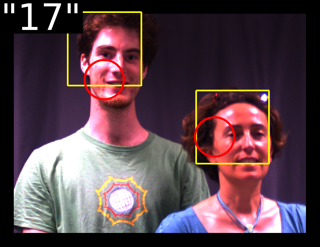}  &
	\includegraphics[width = 0.18\linewidth,clip=,keepaspectratio]{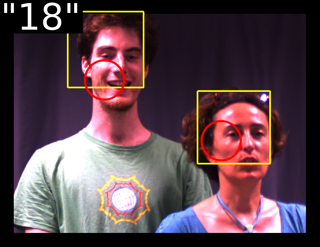}  &
	\includegraphics[width = 0.18\linewidth,clip=,keepaspectratio]{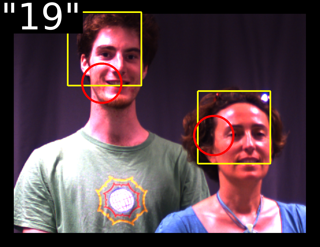}  &
	\includegraphics[width = 0.18\linewidth,clip=,keepaspectratio]{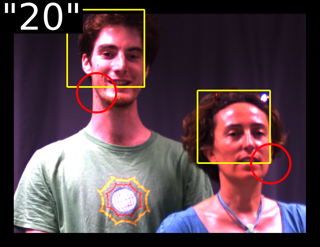}  \\
	\includegraphics[width = 0.18\linewidth,clip=,keepaspectratio]{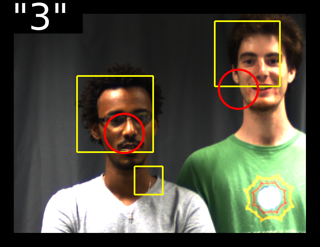}  &
	\includegraphics[width = 0.18\linewidth,clip=,keepaspectratio]{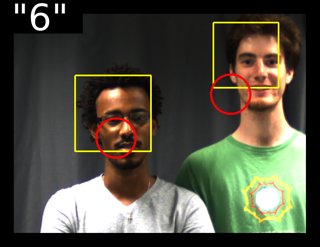}  &
	\includegraphics[width = 0.18\linewidth,clip=,keepaspectratio]{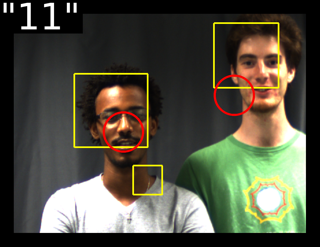}  &
	\includegraphics[width = 0.18\linewidth,clip=,keepaspectratio]{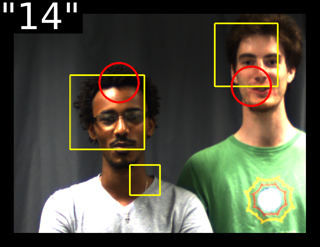}  &
	\includegraphics[width = 0.18\linewidth,clip=,keepaspectratio]{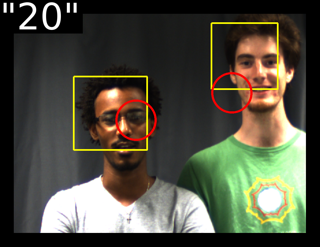} \\
	\includegraphics[width = 0.18\linewidth,clip=,keepaspectratio]{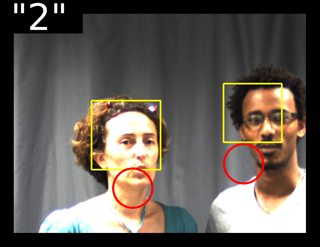} &
	\includegraphics[width = 0.18\linewidth,clip=,keepaspectratio]{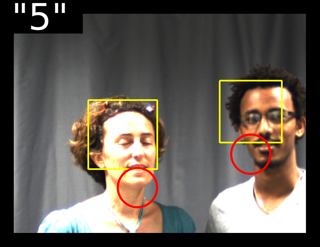} &
	\includegraphics[width = 0.18\linewidth,clip=,keepaspectratio]{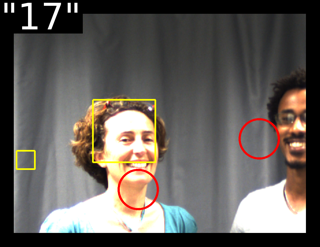} &
	\includegraphics[width = 0.18\linewidth,clip=,keepaspectratio]{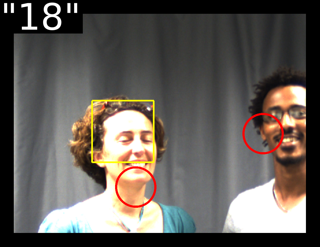} &
	\includegraphics[width = 0.18\linewidth,clip=,keepaspectratio]{twosources_ni5.png} \\
	\includegraphics[width = 0.18\linewidth,clip=,keepaspectratio]{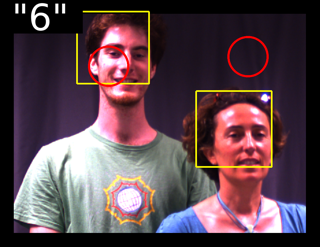} &
	\includegraphics[width = 0.18\linewidth,clip=,keepaspectratio]{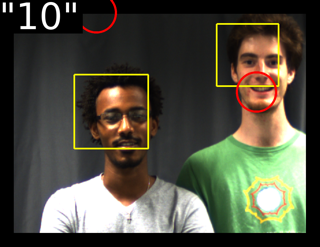} &
	\includegraphics[width = 0.18\linewidth,clip=,keepaspectratio]{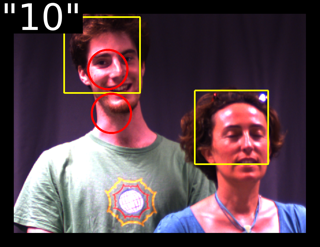} &
	\includegraphics[width = 0.18\linewidth,clip=,keepaspectratio]{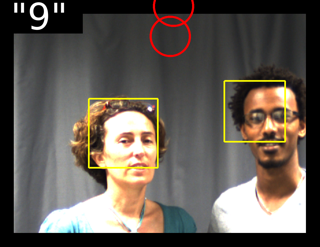} &
	\includegraphics[width = 0.18\linewidth,clip=,keepaspectratio]{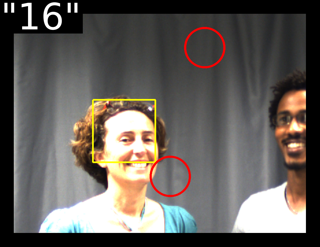} \\
      \end{array}$
      \caption{\label{fig:counting2_scenario} Two subjects count from 1 to 20 (white numbers) in different languages, with a normal voice loudness, from a fixed position. The red circles show the position found with our method. The yellow squares show the results of the Viola-Jones face detector\cite{viola2004robust}. The first 3 rows show examples of successful localization results, the last row shows examples of typical localization errors. Full videos available at https://team.inria.fr/perception/research/binaural-ssl/.}
\end{figure*}
\begin{figure*}
\centering
      \begin{tabular}{cccccc}
	\includegraphics[width = 0.18\linewidth,clip=,keepaspectratio]{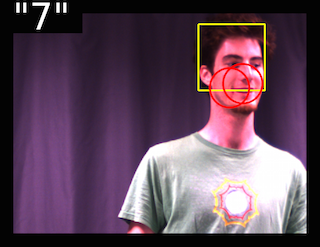}  &
	\includegraphics[width = 0.18\linewidth,clip=,keepaspectratio]{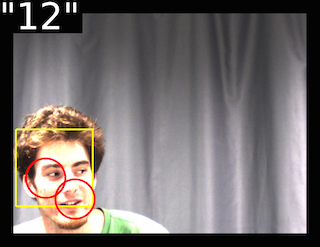}  &
	\includegraphics[width = 0.18\linewidth,clip=,keepaspectratio]{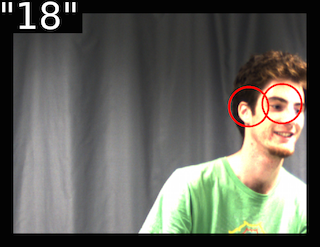}  &
	\includegraphics[width = 0.18\linewidth,clip=,keepaspectratio]{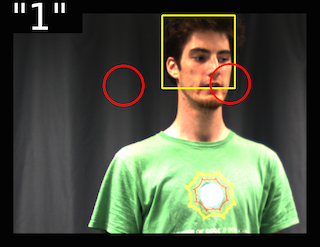}  &
	\includegraphics[width = 0.18\linewidth,clip=,keepaspectratio]{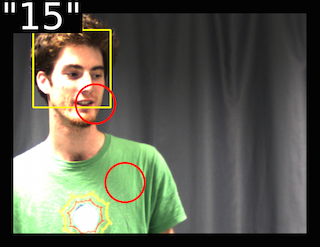}  \\
	(a) & (b) & (c) & (d) & (e)
      \end{tabular}
      \caption{\label{fig:ssl_1source_M2} Examples of results obtained with SBM-2 (red circles) on a single source scenario. The yellow squares are the faces detected with \cite{viola2004robust}. The full video is available at https://team.inria.fr/perception/research/binaural-ssl/.}
\end{figure*}

We further examined the behavior of SBM-2 in two extreme cases. First, we tested the approach on mixtures of two \textit{equal} sound sources, \textit{i.e.}, recordings of two loudspeakers emitting \textit{the same} TIMIT utterance at the same time from two different directions. In that case, the two sources are completely overlapping. Over the 19 test mixtures (38 localization tasks), SBM-2 yielded an average error of $1.5^\circ$ in azimuth, $2.0^\circ$ in elevation, and only one error larger that $10^\circ$. This is similar to results obtained on S+S mixtures with distinct speech signals (Table~\ref{tab:results}). On the other hand, the 3 other methods \cite{Aarabi02,MandelWeissEllis10,DeleforgeForbesHoraud13} failed to localize at least one of the two sources (more than $10^\circ$ error) in more than half of these tests. This result may seem counter-intuitive at first glance. Indeed, a human listener would probably confuse the two identical sources with a single source located somewhere in between the two sources. However, it is in fact unlikely that the set of $D=1536$ frequency-dependent binaural features generated by the two sources matches exactly the set of binaural features generated by a single source at a different location. This result stresses that the proposed co-localization method outperforms traditional WDO-based approaches in heavily overlapping mixtures.

Second, we tested the approach on 100 \textit{non-overlapping} mixtures, \textit{i.e.}, two consecutive 500ms speech segments emitted from different directions. Results obtained with all 4 methods were similar to those obtained for S+S mixtures in Table~\ref{tab:results}. Although ILD and IPD cues depend on the relative spectra of emitting sources (\ref{eq:hrtf_model}), these last experiments show that SBM-2 is quite robust to various type of overlap in mixtures. This is because a large number of binaural features are gathered over the TF plane, thus alleviating perturbations due to varying source spectra (Section \ref{subsec:cues}).

\subsubsection{Person-Live Data}
SBM-2 was also tested on the more realistic \textbf{two-speaker} scenarios. A 1200ms sliding analysis temporal segment was used in order to estimate the two speaker positions. Smaller analysis segments degraded the results. This shows the necessity of gathering enough binaural features for the SBM-2 method to work. Fig.~\ref{fig:counting2_scenario} shows some frames of the video generated from this test. The sound source positions estimated by SBM-2 are shown by a red circle in the corresponding video frame. Three couples of participants were recorded, totaling 124 numbers pronounced. In all experiments, SBM-2 \textit{correctly} localized at least one of the two simultaneous sources in both azimuth and elevation, where correctly means less than $4^\circ$ error (this approximately correspond to the diameter of a face in the image space). Out of the 124 uttered numbers, 92 (74.1$\%$) were correctly localized in both azimuth and elevation. Out of the remaining 32 mistaken localizations, 18 had a correct azimuth estimation (\textit{e.g.}, Fig.~\ref{fig:counting2_scenario} last row, column 1 and 2), 4 were mistaken for the other source (\textit{e.g.}, Fig.~\ref{fig:counting2_scenario} last row, column 3) and only 10 ($8\%$) were incorrectly localized in both azimuth and elevation (\textit{e.g.}, Fig.~\ref{fig:counting2_scenario} last row, column 4 and 5). Results obtained with the Viola-Jones face detection algorithm \cite{viola2004robust} are shown with a yellow square. The face-detector yielded a few erroneous results due to partial occlusions and false detections.

Finally, Fig.~\ref{fig:ssl_1source_M2} shows some examples of the output of SBM-2 on a single source scenario. For 8 out of 20 numbers, the algorithm returned two source positions near the actual source, \textit{e.g.} Fig.~\ref{fig:ssl_1source_M2}(a)(b)(c). This is because the two-source training set also includes mixtures of nearby sources. For the remaining 12 numbers, the algorithm returned one source position near the correct source, and another one at a different location, \textit{e.g.} Fig.~\ref{fig:ssl_1source_M2}(d)(e). This ``ghost'' source may correspond to a reverberation detected by the method.

%% file: conclusionF.tex
\section{Conclusions}
\label{sec:conclusion}
We proposed a supervised approach to the problem of simultaneous localization of audio sources. Unlike existing multiple sound-source localization methods, our method estimates both azimuth and elevation, and requires neither monaural cues segregation nor source separation. Hence, it is intrinsically efficient from a computational point of view. In addition, the approach does not require any camera and/or microphone pre-calibration. Rather, it directly maps sounds onto the image plane, based on a training stage which implicitly captures audio, visual and audio-visual calibration parameters. The proposed method starts by learning an inverse regression between multiple sound directions and binaural features. Then, the learned parameters are used to estimate unknown source directions from an observed binaural spectrogram. Prominently, while the method needs to be trained using a white-noise emitter, it can localize sparse-spectrum sounds, e.g., speech. This is in contrast with other supervised localization methods trained with white-noise. These methods usually localize wide-band sources, or assume that a single source emits during a relatively long period of time, in order to gather sufficient information in each frequency band. This inherently limits their range of practical application scenarios.

We thoroughly tested the proposed method with a new realistic and versatile dataset that is made publicly available. This dataset was recorded using a binaural dummy head and a camera, such that sound directions correspond to pixel locations. This has numerous advantages. First, it provides accurate ground-truth data. Second, it can be used to mix sound samples from existing corpora. Third, the method can then be viewed as an audio-visual alignment method and hence it can be used to associate sounds and visual cues, \textit{e.g.}, aligning speech and faces by jointly using our method and face detection and localization. In the light of our experiments, one may conclude that the proposed method is able to reliably localize up to two simultaneously emitting sound sources in realistic scenarios.

Supervised learning methods for sound-source localization have the advantage that explicit transfer functions are not needed: they are replaced by an implicit representation embedded in the parameters of the regression function. In turn, this requires that the training and testing conditions are similar, \textit{e.g.}, same room and approximatively the same position in the room. In contrast, standard methods assume similarity between simplified transfer function models and real-world conditions. To cope with the limitations of the proposed methods, we plan to train our method over a wider range of source distances, orientations and microphone positions. This could be done in a real room, or alternatively in simulated rooms. Additional latent factors brought by these variations could be captured by adding latent variables to the regression model, as proposed in \cite{DeleforgeForbesHoraud2014b}.

In the future we plan to investigate model selection techniques or to devise a mapping method in the framework of Bayesian non-parametric models, in order to automatically select the number of sources. We will also scale up the method to more than 2 sources, using parallelization in the training stage, and increase the number of microphones. Finally, to reduce the number of false detections in live experiments, we plan to use a more advanced speech activity detector to automatically adjust the analysis window, and to use a tracker to take into account past observations. This could be done by incorporating a hidden Markov chain to our probabilistic model.

%% file: biographies.tex
\begin{IEEEbiography}[{\includegraphics[width=1in,
height=1.25in,clip,keepaspectratio]{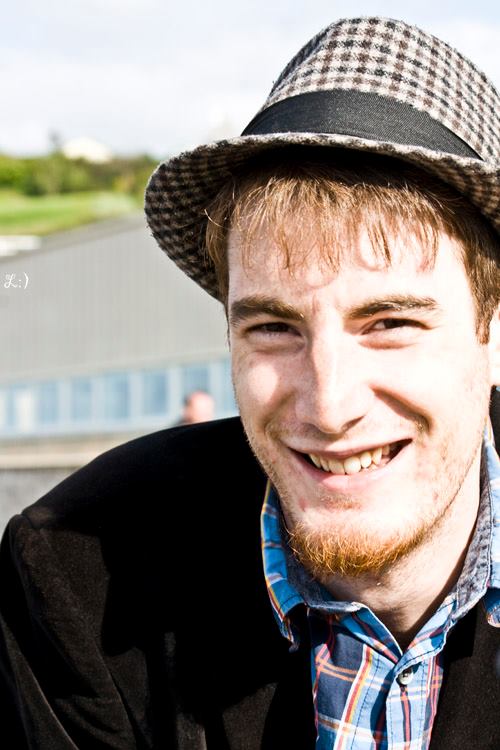}}]{Antoine Deleforge}
received the B.Sc. (2008) and M.Sc. (2010) engineering degrees in computer science and mathematics from the Ensimag engineering school (Grenoble, France) as well as the specialized M.Sc. (2010) research degree in computer graphics, vision, robotics from the Universit\'e Joseph Fourier (Grenoble, France). In 2013, he received the Ph.D. degree in computer science and applied mathematics from the university of Grenoble (France). He is employed since January 2014 as a post-doctoral fellow at the chair of Multimedia Communication and Signal Processing of the Erlangen-Nuremberg University (Germany). His research interests include machine learning for signal processing, Bayesian statistics, computational auditory scene analysis, and robot audition.
\end{IEEEbiography}

\begin{IEEEbiography}[{\includegraphics[width=1in,
height=1.25in,clip,keepaspectratio]{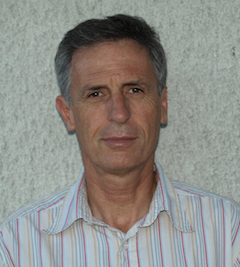}}]{Radu Horaud} 
received the B.Sc. degree in electrical engineering, the M.Sc. degree
in control engineering, and the Ph.D. degree in computer science from
the Institut National Polytechnique de Grenoble, Grenoble, France. 
Currently he holds a position of director of research with the Institut National de Recherche en Informatique et
Automatique (INRIA), Grenoble Rh\^one-Alpes, Montbonnot, France, where
he is the founder and head of the PERCEPTION team. His
research interests include computer vision, machine learning, audio signal processing, 
audiovisual analysis, and robotics. He is an area editor of the
\textit{Elsevier Computer Vision and Image Understanding}, a member of
the advisory board of the \textit{Sage International Journal of Robotics
  Research}, and an associate editor of the
\textit{Kluwer International Journal of Computer Vision}. He was
Program Cochair of the Eighth IEEE International Conference on
Computer Vision (ICCV 2001). In 2013, Radu Horaud was awarded a five year ERC Advanced Grant for his project \textit{Vision and Hearing in Action} (VHIA).
\end{IEEEbiography}

\begin{IEEEbiography}[{\includegraphics[width=1in,
height=1.25in,clip,keepaspectratio]{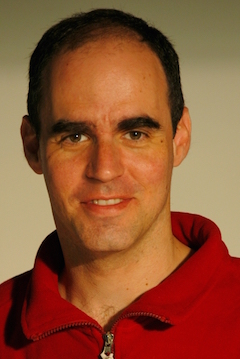}}]{Yoav Y. Schechner}
received his BA and MSc degrees in physics and PhD in electrical engineering from the Technion-Israel Institute of Technology in 1990,1996, and 2000, respectively. During the years 2000 to 2002 Yoav was a research scientist at the computer science department in Columbia University. Since 2002, he is a faculty member at the department of Electrical Engineering of the Technion, where he heads the Hybrid Imaging Lab. From 2010 to 2011 he was a visiting Scientist in
Caltech and NASA's Jet Propulsion Laboratory. His research is focused on computer vision, the use of optics and physics in imaging and computer vision, and on multi-modal sensing. He was the recipient of the Wolf Foundation Award for Graduate Students in 1994, the Guttwirth Special Distinction Fellowship in 1995, the Ollendorff Award in 1998, the Morin Fellowship in 2000- 2002, the Landau Fellowship in 2002-2004 and the Alon Fellowship in 2002-2005. He has received the Klein Research Award in 2006, the Outstanding Reviewer Awards in IEEE CVPR 2007 and ICCV 2007 and the Best Paper Award in IEEE ICCP 2013.
\end{IEEEbiography}

\begin{IEEEbiography}[{\includegraphics[width=1in,
height=1.25in,clip,keepaspectratio]{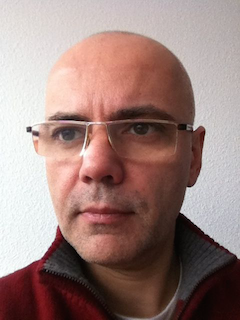}}]{Laurent Girin}
received the M.Sc. and Ph.D. degrees in signal processing from the Institut National Polytechnique de Grenoble (INPG), Grenoble, France, in 1994 and 1997, respectively. In 1999, he joined the Ecole Nationale Supe ́rieure d’Electronique et de Radioe ́lectricite ́ de Grenoble (ENSERG), as an Associate Professor. He is now a Professor at Phelma (Physics, Electronics, and Materials Department of Grenoble-INP), where he lectures signal processing theory and applications to audio. His research activity is carried out at GIPSA-Lab (Grenoble Laboratory of Image, Speech, Signal, and Automation). It deals with different aspects of speech and audio processing (analysis, modeling, coding, transformation, synthesis), with a special interest in joint audio/visual speech processing and source separation. Prof. Girin is also a regular collaborator at INRIA (French Computer Science Research Institute), Grenoble, as an associate member of the Perception Team. 
\end{IEEEbiography}

\vfill